\documentclass[useAMS,usenatbib,usegraphicx,onecolumn]{mn2e}

\usepackage{amssymb,amsfonts,amsmath,amstext,amsgen,amsopn,amsxtra,indentfirst,lscape,times}

\title[Vortices as Nurseries for Planetesimal Formation]{Vortices as nurseries for planetesimal formation \\ in protoplanetary discs}
\author[Heng \& Kenyon]{Kevin Heng$^{1,2}$\thanks{E-mail:
heng@ias.edu (KH); skenyon@cfa.harvard.edu (SK)} and Scott J. Kenyon$^{3}$\footnotemark[1]\\
$^{1}$Frank \& Peggy Taplin Member, Institute for Advanced Study, School of Natural Sciences, Einstein Drive, Princeton, NJ 08540, U.S.A.\\
$^{2}$Zwicky Fellow, ETH Z\"{u}rich, Institute for Astronomy, Wolfgang-Pauli-Strasse 27, CH-8093, Z\"{u}rich, Switzerland\\
$^{3}$Harvard-Smithsonian Center for Astrophysics, 60 Garden Street, Cambridge, MA 02138, U.S.A.}

\begin{document}

\date{Submitted 2010 May 8.  Re-submitted 2010 June 10.  Accepted 2010 June 15.}

\pagerange{\pageref{firstpage}--\pageref{lastpage}} \pubyear{2010}

\maketitle

\label{firstpage}

\begin{abstract}
Turbulent, two-dimensional, hydrodynamic flows are characterized by the emergence of coherent, long-lived vortices without a need to invoke special initial conditions.  Vortices have the ability to sequester particles, with typical radii $\sim 1$ mm to $\sim 10$ cm, that are slightly decoupled from the gas.  A generic feature of discs with surface density and effective temperature profiles that are decreasing, power-law functions of radial distance is that four vortex zones exist for a fixed particle size.  In particular, two of the zones form an annulus at intermediate radial distances within which small particles reside.  Particle capture by vortices occurs on a dynamical time scale near and at the boundaries of this annulus.  As the disc ages and the particles grow via coagulation, the size of the annulus shrinks.  Older discs prefer to capture smaller particles because the gas surface density decreases with time, a phenomenon we term ``vortex aging''.  More viscous, more dust-opaque and/or less massive discs can have vortices that age faster and trap a broader range of particle sizes throughout the lifetime of the disc.  Thus, how efficiently a disc retains its mass in solids depends on the relative time scales between coagulation and vortex aging.  If vortices form in protoplanetary discs, they are important in discs with typical masses and for particles that are likely to condense out of the protostellar nebula.  Particle capture also occurs at distances relevant to planet formation.  Future infrared, submillimetre and centimetre observations of grain opacity as a function of radial distance will test the hypothesis that vortices serve as nurseries for particle growth in protoplanetary discs.
\end{abstract}


\section{Introduction}

The formation of $\sim$ km-sized planetesimals from sub-micron-sized dust grains likely involves more than one physical process \citep{youdin10,cy10}.  It is generally accepted that particle growth in dusty, circumstellar discs is hierarchical, eventually forming planetesimals that are the building blocks of planets \citep{armitage07,armitage10}.  There is reasonable understanding of building small particles, but forming planetesimals from these particles remains mired in controversy.  The initial stage of growth probably proceeds through the nucleation of sub-micron-sized dust grains from the primordial nebula, which then form the monomers of fractal dust aggregates up to $\sim 1$ mm to $\sim 10$ cm sizes in $\gtrsim 10^3$ yrs, beyond which growth is stalled by collisional bouncing and fragmentation \citep{bw08,zsom10}.  In this regime, the particle dynamics and coagulation are described by Brownian motion and van der Waals forces.  One of the best astrophysical pieces of evidence for grain growth to these sizes is the detection of 3.5 cm dust emission from the classical T Tauri star TW Hya (age $\sim 5$--10 Myr), located 56 pc away, which has a face-on circumstellar disc of radius 225 AU \citep{wilner05}.

In standard models of protoplanetary discs, the gas pressure decreases radially outward.  Gas in the disc then moves at sub-Keplerian speeds.  Solid particles on Keplerian orbits experience a ``head wind'' with a velocity $\sim 10^3$ cm s$^{-1}$ --- this head wind drags $\sim 10$ cm- to $\sim 1$ m-sized particles into the central star on time scales of $\sim 10$--100 years \citep{weiden77a}.  These time scales are much shorter than the characteristic time scale for these particles to collide and grow into larger particles that are unaffected by the head wind.  \cite{safronov69} and \cite{gw73} suggested that this difficulty can be circumvented by dust settling into the midplane of the disc and triggering gravitational instability, but \cite{weiden80} pointed out that turbulence generated by the settling impedes the process.  

Long-lived structures in the gas are a possible way of concentrating particles with sizes $\sim 1$ mm to $\sim 10$ cm and growing them to larger sizes.  Such structures are usually high-pressure regions in the gas that are capable of concentrating particles, which is a manifestation of Bernoulli's principle\footnote{Bernoulli's principle states that in inviscid flows, a decrease in fluid velocity is accompanied by an increase in pressure.  Hence, locations of maximum pressure are also locations of minimum velocity.} \citep{kundu04}.  Vortices are examples of long-lived structures --- they are spiral, non-linear motions of fluid with closed streamlines.  On Earth, ocean vortices have been observed to trap larval fish off the coast of western Australia \citep{paterson08}.  In astrophysical settings, the possible role of vortices in planet formation was suggested by \cite{vw46}, based on the writings of Kant, in an article entitled {\it Die Entstehung des Planetensystems} (``The Origin of Planetary Systems'').  Since then, vortices have been suggested as possible nurseries for growth to $\gtrsim 1$ m-sized particles \citep{aw95,bs95,tanga96,bracco99,gl99,gl00,johansen04,bm05,fn05,kb06,inaba06,shen06,bodo07,mr09}.  In stratified protoplanetary discs, there are conceivable locations where the flow is turbulent and quasi-2D.  An attractive feature of turbulent, 2D, hydrodynamic flows is that the fluid robustly self-organizes into large, coherent, long-lived vortices amidst a backdrop of small eddies without a need to invoke special initial conditions \citep{carn91,wm93,tabeling02}.  Such a property arises from the fact that the so-called ``vortex-stretching term'' in the vorticity equation is absent in 2D, thereby allowing an inverse cascade of energy (and forward cascade of enstrophy).  Turbulence thus becomes a friend and not a foe.  If protoplanetary discs are capable of producing turbulent, quasi-2D flows, then these may seed large-scale vortices that may survive for many orbital time scales.  Such optimism should be tempered by the fact that off-midplane vortices have not been observed in simulations of protoplanetary discs with dust settling \citep{chiang08,johansen09}.

\emph{The main question we are addressing in this study is: assuming vortices can be generated and sustained in discs, what sizes of particles do they capture, where are the capture locations and when does capture occurs?}  In \S\ref{sect:basic}, we discuss/review the order-of-magnitude physics associated with protoplanetary discs, particle-gas interactions and vortices.  In \S\ref{sect:static}, we start with the simpliest case of a static, minimum mass solar nebula disc and show that there are generically four vortex zones within any disc with surface density and temperature profiles that are decreasing, power-law functions of $r$.  In \S\ref{sect:viscous}, we consider the next level of sophistication, which is the case of an evolving, viscously-heated disc; we show that there are preferred locations and particle radii for vortex capture.  In \S\ref{sect:irradiated}, we generalize to the case of an evolving, viscous, irradiated disc.  We demonstrate that discs which are able to both settle particles to their midplanes and capture them via vortices have upper limits to their masses that are consistent with most observed discs.  We also show that the maximum particle radius for vortex capture in these discs is $\sim 10$ cm, independent of disc model and weakly dependent on stellar, disc and dust properties.  In \S\ref{sect:discussion}, we summarize our conclusions, discuss the open questions concerning the physics of vortices and describe the relevance of our results to observations.  Table \ref{tab:parameters} lists the fiducial values adopted for the parameters of our models.

\section{Order-of-Magnitude Physics: Discs, Particle-Gas Interactions and Vortices}
\label{sect:basic}

\begin{table*}
\centering
\caption{Fiducial parameter values}
\label{tab:parameters}
\begin{tabular}{lcc}
\hline\hline
\multicolumn{1}{c}{Parameter(s)} & Description & \multicolumn{1}{c}{Adopted value(s)}\\
\hline
\vspace{2pt}
$M_0$ & initial disc mass & 0.01 $M_\odot$ \\
$s_0$ & initial disc outer radius & 20 AU \\
$M_\star, T_\star, R_\star$ & stellar mass, temperature, radius & $M_\odot, T_\odot, R_\odot$ \\
$\rho_s$ & material/internal density & 3 g cm$^{-3}$ \\
$\kappa_0$ & dust opacity & 1 cm$^2$ g$^{-1}$ \\
$\alpha$ & viscosity parameter & 0.01 \\
$\gamma$ & gas adiabatic index & 1.4 \\
$\mu$ & mean molecular weight of gas & 2.4 \\
\hline
\hline
\end{tabular}\\
\end{table*}

We consider vortices in an axisymmetric, viscous accretion disc with surface density 
$\Sigma(r,t)$ and (effective) temperature $T(r,t)$, where $r$ is the radial distance from the star and $t$
is the time.  Disc material orbits with angular velocity $\Omega(r) = (G M_\star / r^3 )^{1/2}$ 
around a star with mass $M_\star$.  The sound speed in a disc depends on $T(r,t)$ and is typically
\begin{equation}
c_s = \sqrt{\frac{\gamma k_{\rm B}T}{m_{\rm H}}} \approx 1 \mbox{ km s}^{-1} ~\gamma^{1/2} \left(\frac{T}{100 \mbox{ K}}\right)^{1/2},
\label{eq:cs}
\end{equation}
where $\gamma$ is the adiabatic index of the gas, $k_{\rm B}$ is the Boltzmann constant, $m_{\rm H}$ is the mass of a hydrogen atom and we have ignored the radial dependence of temperature within the disc for simplicity.  In the vertical direction, the gas is in hydrostatic equilibrium with vertical scale height 
\begin{equation}
H = \frac{c_s}{\Omega} \approx 0.03 \mbox{ AU} ~\left(\frac{T}{100 \mbox{ K}}\right)^{1/2} \left(\frac{M_\star}{M_\odot}\right)^{-1/2} \left(\frac{r}{\mbox{AU}}\right)^{3/2}.
\label{eq:h}
\end{equation}
Gas diffuses inward at a rate $\dot{M} = 3 \pi \nu \Sigma$, where $\nu$ is the viscosity. We adopt an ``alpha-model'' for the viscosity \citep{ss73}, where
\begin{equation}
\nu = \alpha c_s H = \frac{\alpha c^2_s}{\Omega}.
\label{eq:nu}
\end{equation}

There are three important time scales associated with the disc. The shortest is the dynamical (orbital) time scale,
\begin{equation}
t_d \sim \Omega^{-1} \approx 0.2 \mbox{ yr} ~\left(\frac{M_\star}{M_\odot} \right)^{-1/2} \left(\frac{r}{\mbox{AU}} \right)^{3/2}.
\end{equation}
The disc cools on the local thermal timescale \citep{pri1981}, which is the ratio of the thermal 
energy of the gas ($\Sigma c_s^2$) to the rate of viscous energy dissipation ($\nu \Sigma \Omega^2$).
Using our expression for $\nu$ (equation [\ref{eq:nu}]), the cooling time is
\begin{equation}
t_{\rm cool} \sim \alpha^{-1} t_d \approx 20 \mbox{ yr} ~\left(\frac{\alpha}{0.01} \right)^{-1} \left(\frac{M_\star}{M_\odot} \right)^{-1/2} \left(\frac{r}{\mbox{AU}} \right)^{3/2}.
\label{eq:t_c}
\end{equation}
Thus, the disc cools on time scales much longer than the dynamical time.  The viscous time scale is usually the longest characteristic time and measures the rate at which matter diffuses through the disc,
\begin{equation}
t_{\rm vis} \sim \frac{r^2}{\nu} \sim \alpha^{-1} \left(\frac{r}{H} \right)^2 t_d \sim  \left(\frac{r}{H} \right)^2 t_{\rm cool} \approx 2 \times 10^5 \mbox{ yr} ~\left(\frac{r/H}{100} \right)^2 \left(\frac{\alpha}{0.01} \right)^{-1} \left(\frac{M_\star}{M_\odot} \right)^{-1/2} \left(\frac{r}{\mbox{AU}} \right)^{3/2}.
\label{eq:vis_0}
\end{equation}
It is apparent that $t_{\rm vis} \gg t_{\rm cool} \gg t_d$ in typical discs \citep{lbp74,pri1981}.

In a Keplerian disc, solving the radial momentum equation yields the velocity difference between a particle and its surrounding gas \citep{pri1981}. For a thin disc, $c_s \ll v_{\rm K}$ where $v_{\rm K} = (GM_\star/r)^{1/2}$. Defining inward radial drift to have a positive sign,
\begin{equation}
\Delta v \approx \left(\frac{p}{2} + \frac{q}{4} + \frac{3}{4} \right) \frac{c^2_s}{v_{\rm K}}.
\label{eq:dv}
\end{equation}
The quantities $p$ and $q$ are the power-law indices of the surface density and temperature profiles as functions of $r$,
\begin{equation}
p = -\frac{\partial \ln{\Sigma}}{\partial \ln{r}}, ~q = -\frac{\partial \ln{T}}{\partial \ln{r}}.
\end{equation}
From Equation (\ref{eq:dv}), a pressure gradient that declines radially outward produces an inward radial drift of particles.  If $p$ and $q$ are sufficiently negative, then the particles do not drift.  If vortices can provide this kind of environment, they will trap particles.  Ignoring the dependence of $\Delta v$ on $p$ and $q$, we have
\begin{equation}
\frac{\Delta v}{c_s} \sim \frac{c_s}{v_{\rm K}} \approx 0.03 ~\left(\frac{T}{100 \mbox{ K}} \frac{r}{\mbox{AU}}\right)^{1/2} \left(\frac{M_\star}{M_\odot}\right)^{-1/2}.
\end{equation}
For $c_s \sim$ 1 km s$^{-1}$ (equation [\ref{eq:cs}]), typical drift speeds are roughly 30 m s$^{-1}$. \citet{adachi76} and \citet{weiden77a} derived detailed expressions for the time scale of radial drift $t_{\rm drift}$. As a rough estimate, $t_{\rm drift} \sim r/\Delta v \sim 10^2$ yr $(r/\mbox{AU})$, longer than the dynamical or cooling timescales but much shorter than the viscous timescale or the expected dissipation time of the nebular gas ($\sim 10^6$--$10^7$ yr).  Computing $t_{\rm drift}$ more accurately requires a careful consideration of the drag coefficients, which themselves depend on $a$ \citep{adachi76,weiden77a}.  To facilitate discussion, we estimate an approximate radial drift time scale at $r \sim 1$ AU as
\begin{equation}
t_{\rm drift} \sim 10 \mbox{ yr}
\begin{cases}
\left( a / 1 \mbox{ m} \right)^{-1}, & a < 1 \mbox{ m},\\
a / 1 \mbox{ m}, & a > 1 \mbox{ m}.\\
\end{cases}
\end{equation}
In the absence of turbulence, the time scale for particles to settle to the midplane of the disc is \citep{cg97}
\begin{equation}
t_{\rm settle} \sim \frac{\Sigma}{a \rho_s \Omega} \approx 500 \mbox{ yr} ~\left(\frac{\Sigma}{10^3 \mbox{ g cm}^{-2}}\right) \left(\frac{a}{1 \mbox{ mm}} \frac{\rho_s}{3 \mbox{ g cm}^{-3}}\right)^{-1} \left(\frac{M_\star}{M_\odot} \right)^{-1/2} \left(\frac{r}{\mbox{AU}} \right)^{3/2},
\label{eq:settle}
\end{equation}
where $\rho_s$ is the material/internal density of the particles and we have ignored the radial dependence of the surface density.  Equation (\ref{eq:settle}) is effectively the \emph{minimum} time for the particles to settle, as vertical mixing may hold the particles aloft.  Even turbulence as weak as $\alpha \sim 10^{-9}$ may frustrate the settling process (see \citealt{cuzzi08} and references therein).  (If the particles are porous/fractal, the settling time may also increase.)  Particles with radii $a \ll 1$ mm ($t_{\rm drift} > 10^4$ yr) are well-coupled to the gas and settle somewhat slowly to the disc midplane.  Particles with radii $a \sim 1$ mm to $\sim 10$ cm ($t_{\rm drift} \sim 10^2$--$10^4$ yr) are weakly-coupled and settle on time scales $\gtrsim 5$--$500$ yr. The largest particles with radii $\gg 1$ m are uncoupled and do not ``see'' the gas.  Thus, particles are considered ``small'' or ``large'' depending on the extent to which they are coupled to the gas, a concept we will develop later in the context of vortices.  Small particles may grow up to some maximum particle size, typically $\lesssim 10$ cm, before starting to drift inward.  We wish to show that vortices are capable of capturing particles in this size range as they drift radially inward through the disc.
 
The typical growth time for small particles is shorter than the drift time.  Particles larger than micron sizes are probably fractal aggregate structures (see Appendix \ref{append:coagulate}), which simulations and laboratory experiments suggest take \citep{bw08,zsom10}
\begin{equation}
t_{\rm coag} \gtrsim 10^3 \mbox{ yr}
\end{equation}
to form. Once these particles form and start to settle out of the gas, the collisional time scale for an ensemble of solid particles is \citep{ht10},
\begin{equation}
t_c \sim \frac{a \rho_s}{\Sigma_s \Omega} \approx 5 \times 10^{-3} \mbox{ yr} ~\left(\frac{a}{1 \mbox{ mm}} \frac{\rho_s}{3 \mbox{ g cm}^{-3}}\right) \left(\frac{\Sigma_s}{10 \mbox{ g cm}^{-2}}\right)^{-1} \left(\frac{M_\star}{M_\odot} \right)^{-1/2} \left(\frac{r}{\mbox{AU}} \right)^{3/2},
\end{equation}
where $\Sigma_s$ is the surface density of solids (typically 1\% of $\Sigma$) and we have ignored gravitational focusing.  If collisions result in the growth to larger particles, then particles with $a \sim 1$ mm to $\sim 10$ cm grow more rapidly ($t_c \approx 0.005$--$0.5$ yr) than they drift through the gas ($t_{\rm drift} \sim 10^4$--$10^2$ yr).  Larger particles with $a \approx$ 1 m grow on timescales ($\sim 5$ yr) comparable to their drift times. Thus, producing particles much larger than 1 m requires an environment to concentrate small particles in regions of larger local surface density where they can grow into much larger planetesimals which are safe from radial drift.  Vortices provide this environment.

Particles in the vicinity of a vortex experience centrifugal (due to vortex and not disc rotation) and Coriolis forces, directed outward and inward respectively \citep{chav00}.  In polar coordinates centered on the vortex $(R,\theta)$, the relevant terms in the equation for the radial acceleration ($d^2R/dt^2$) of the particle are:
\begin{equation}
\begin{split}
&\mbox{Centrifugal: } ~R \left( \frac{d\theta}{dt} \right)^2,\\
&\mbox{Coriolis: } ~2 \Omega R \frac{d\theta}{dt}.\\
\end{split}
\end{equation}
Only anti-cyclonic vortices ($d\theta/dt < 0$) direct particles toward the vortex centers; they also survive longer than cyclonic vortices \citep{davis00}.  For net inward acceleration of the particle to exist, we must have $\vert d\theta/dt \vert < 2 \Omega$, a condition which is always fulfilled in practice.  Two processes --- viscous dissipation and orbital shear --- limit the sizes of vortices. Viscous dissipation destroys vortices smaller than the viscous length scale, 
\begin{equation}
L_{\rm vis} = \frac{\alpha c_s H}{v_{\rm vor}} \approx 0.003 \mbox{ AU} ~\left( \frac{\alpha}{0.01} \frac{T}{100 \mbox{ K}} \right) \left( \frac{v_{\rm vor}}{0.1 ~c_s} \right)^{-1} \left(\frac{M_\star}{M_\odot}\right)^{-1/2} \left(\frac{r}{\mbox{AU}}\right)^{3/2},
\end{equation}
on time scales comparable to $t_{\rm vis}$ (equation [\ref{eq:vis_0}]), where $v_{\rm vor} \sim 0.1 ~c_s$ \citep{inaba06} is the rotational speed of the vortex.  Even in this case, vortices formed at this scale are many orders of magnitude larger than the particles considered and can survive for many dynamical times.  Keplerian shear inhibits the formation of circular structures larger than the shear length scale \citep{gl99},
\begin{equation}
L_{\rm shear} = \sqrt{v_{\rm vor} \left\vert \frac{\partial \Omega}{\partial r} \right\vert^{-1}} \approx 0.05 \mbox{ AU} ~\left( \frac{v_{\rm vor}}{0.1 ~c_s} \right)^{1/2} \left(\frac{M_\star}{M_\odot}\right)^{-1/4} \left(\frac{r}{\mbox{AU}}\right)^{5/4}.
\end{equation}
Circular vortices forming at scales $\gtrsim L_{\rm shear}$ get elongated in the azimuthal direction, which allows them to survive longer.  These estimates for $L_{\rm vis}$ and $L_{\rm shear}$ show that vortices are large-scale phenomena with length scales $L \gg a$.

In 3D, vortices are generally subjected to hydrodynamic instabilities that may destroy them on time scales shorter than $t_{\rm vis}$ (see \S\ref{sect:discussion}).  In 2D, vortices may live for many dynamical times due to the absence of the vortex trapping term in the vorticity equation.  An inverse energy cascade exists in 2D turbulence, allowing large-scale vortices to emerge naturally from self-organization of the fluid \citep{tabeling02}.  It is therefore much easier to make statements about vortices in 2D.  The figure of merit for whether a fluid flow is 2D or 3D is the Froude number, which is a measure of how rapidly a fluid element responds, against the vortex flow, when vertically displaced in a convectively stable fluid layer.  At a given vertical height $z$ in the disc, the fluid layer is effectively 2D if the Froude number is less than unity \citep{bm05}.  It depends on the ratio of two quantities: the buoyancy time scale $t_{\rm BV}$ and the rotational period of the gas around the vortex $t_{\rm vor}$.  Assuming vertical isothermality for simplicity, we get
\begin{equation}
t_{\rm BV} \sim \frac{2\pi}{\Omega} \left(\frac{H}{z} \right).
\end{equation}
As reasoned by \cite{bm05}, the Keplerian shear is comparable in magnitude to the rotational rate, implying that $t_{\rm vor} \sim 2\pi/\Omega$.  The Froude number is then
\begin{equation}
{\cal F} = \frac{t_{\rm BV}}{t_{\rm vor}} \sim \frac{H}{z},
\label{eq:froude}
\end{equation}
where we have ignored a factor of order unity related to the aspect ratio of the vortex.  Near the disc mid-plane, ${\cal F} \gg 1$ and the flow is 3D.  A couple of vertical scale heights or more from the mid-plane, however, the flow already becomes quasi-2D --- any turbulence generated in these fluid layers will inevitably seed vortices, but we again note that such off-midplane vortices are not seen in simulations \citep{chiang08,johansen09}.

\cite{chav00} considered the forces acting on a particle crossing a vortex.  By seeking solutions of the form,
\begin{equation}
\begin{split}
&x = L \cos{\left(\omega t\right)} \exp{\left(-t/t_{\rm cap}\right)},\\
&y = L^\prime \sin{\left(\omega t\right)} \exp{\left(-t/t_{\rm cap}\right)},\\
\end{split}
\end{equation}
where $(x,y)$ are the Cartesian coordinates of the particle with the origin centered on the vortex, he was able to show that $L^\prime = L/\chi$ and $\omega = -3\Omega/2(\chi-1)$ --- the particles follow ellipses of aspect ratio $\chi \ge 1$ and move with angular velocity $\omega$.  The eccentricity of the ellipse is $(\chi^2-1)^{1/2}/\chi$.  The combined effect of the Coriolis force and drag make the particles drift towards the vortex center on a time scale:
\begin{equation}
t_{\rm cap} = 
\begin{cases}
\frac{4\chi \left(\chi-1\right)^2}{3 \left(\chi-2\right) \left(2\chi+1\right)} \left(\frac{\xi}{\Omega^2}\right), & \xi > \Omega, \\
\frac{2\chi \left(\chi-1\right)}{\left(\chi-3\right) \left(2\chi+1\right)} \left(\frac{1}{\xi}\right), & \xi < \Omega, \\
\end{cases}
\label{eq:tcap}
\end{equation}
where $\xi$ is the friction coefficient which we will describe shortly.  The preceding expressions were derived by \cite{chav00} under the assumption that the particle motion within the vortex is deterministic, which becomes invalid if the vortex cores become strongly turbulent \citep{lp10}.  Since $\chi=4$ minimizes the capture time \citep{chav00}, we will adopt this value for the rest of the paper whenever necessary and note that the numerical coefficients in equation (\ref{eq:tcap}) are both 8/3.  The capture time depends on particle size and the properties of the gas through the friction coefficient $\xi$.  Denoting the mean free path of the gas as $\lambda \sim 1$ cm, particles are in the Epstein (Stokes) regime when $a < 9 \lambda /4$ ($a > 9 \lambda /4$) \citep[e.g.,][]{adachi76, weiden77a}.  For ``small'' particles coupled closely to the gas, the transition from the Epstein regime to the Stokes regime occurs at a disc radius $r = r_c$ (see \S3). Thus, we write the friction coefficient as \citep{weiden77a,cuzzi93},
\begin{equation}
\xi = 
\begin{cases}
\frac{\Sigma \Omega}{2 \rho_s a}, & \mbox{ Epstein regime } \left(a < 9\lambda/4 \mbox{ or } r > r_{\rm c} \right), \\
\frac{9 \Sigma \Omega \lambda}{8 \rho_s a^2}, & \mbox{ Stokes regime } \left(a > 9\lambda/4 \mbox{ or } r < r_{\rm c} \right). \\
\end{cases}
\end{equation}
Solving for the vortex structure within the disc then requires specifying a disc model that in turn specifies $\Sigma$, $T$ and $\lambda$ as functions of $r$ and/or $t$, which we will explore in the subsequent sections (\S\ref{sect:static}, \S\ref{sect:viscous} and \S\ref{sect:irradiated}).  It is worth noting that $\Omega/\xi$ is the commonly-used ``Stokes number''.  If $\xi/\Omega \gg 1$, the particle is small and tightly coupled to the gas, implying that it is unable to be trapped at the centre of the vortex.  If $\xi/\Omega \ll 1$, the particle is large and does not ``see'' the vortex.  Hence, as already noted by \cite{chav00} and simulated by \cite{johansen04}, vortices tend to ``pick out'' particles of a certain size, i.e., $\xi/\Omega \sim 1$, such that capture occurs within an orbital period, $t_{\rm cap} \sim \Omega^{-1}$.  These particles are also the ones that attempt to settle towards the mid-plane of the disc after growing to large sizes via coagulation.  This aerodynamic sorting of the particles inside vortices is consistent with evidence that meteorites in our Solar System are composed of chondrules\footnote{Chondrules are typically mm-sized particles found in chondrites, which are meteorites with near-solar compositions unaltered by heating processes.} of a similar size \citep{hewins97}.  It is worthwhile to note that the expression for $\xi$ in the Stokes regime only holds when the particle Reynolds number is less than unity.  Equivalently, this requires
\begin{equation}
\frac{c_s}{v_{\rm K}} < \frac{1}{3+q+2p} \left(\frac{\lambda}{a} \right).
\label{eq:stokes_condition}
\end{equation}
Since $c_s/v_{\rm K} \ll 1$ for the discs we are considering, we expect the condition in equation (\ref{eq:stokes_condition}) to be satisified.  

Finally, gravitational instability may be triggered if the concentrated mass density within the vortices exceeds the \emph{effective} Roche density,
\begin{equation}
\rho_{\rm R} = \frac{3 \varpi M_\star}{4\pi r^3} \sim 10^{-7} \mbox{ g cm}^{-3} ~\varpi ~\left(\frac{M_\star}{M_\odot} \right) \left(\frac{r}{\mbox{AU}} \right)^{-3},
\end{equation}
where $\varpi$ is a dimensionless factor accounting for the delay of gravitational collapse due to gas pressure.  Setting $\varpi=1$ gives the traditional mass density threshold for gravitational instability, which is about two orders of magnitude larger than the characteristic mass density of the gas within the nebula, $\rho_0 \sim 10^{-9}$ g cm$^{-3}$.  Traditionally, gravitational collapse occurs within
\begin{equation}
t_{\rm G} \sim \frac{\pi}{\left(G \rho_{\rm c} \right)^{1/2}} \approx 1 \mbox{ yr } ~\left(\frac{\rho_{\rm c}}{10^{-7} \mbox{ g cm}^{-3}} \right)^{-1/2},
\end{equation}
where $\rho_{\rm c} \gtrsim \rho_{\rm R}$ is the mass density of the self-gravitating clump created by gravitational instability.  Among others, \cite{cuzzi08} point out that gas pressure and turbulence act to delay the onset of collapse, analogous to the role of ambipolar diffusion in star formation, estimate $\varpi \gtrsim 10^2$ and show instead that gravitational collapse occurs on a timescale,
\begin{equation}
t_{\rm G,eff} \sim \frac{1}{4\pi^2} \left(\frac{\rho_0 c_s}{\rho_s a} \right) t^2_{\rm G} \approx 400 \mbox{ yr} ~\left(\frac{\rho_0}{10^{-9} \mbox{ g cm}^{-3}} \frac{c_s}{1 \mbox{ km s}^{-1}} \right) \left(\frac{\rho_{\rm c}}{10^{-7} \mbox{ g cm}^{-3}} \frac{\rho_s}{3 \mbox{ g cm}^{-3}} \frac{a}{1 \mbox{ mm}}\right)^{-1}.
\end{equation}

In summary, small particles ($\lesssim 1~\mu$m), strongly coupled to the gas, coagulate rapidly and attempt to settle to the midplane of the disc.  Particles approaching $\sim 1$ mm to $\sim 10$ cm in size experience radial drag forces that direct them into the star.  Vortex capture occurs on a dynamical time scale --- large-scale vortices therefore provide a way to trap these particles before they drift into the star.  The enriched particle density within the vortices then enhances particle growth \citep{inaba06}.  Growing particles remain trapped within the vortices until they are large enough to decouple from the gas, when radial drag is negligible.  Thus, vortices provide a natural way for small particles to grow into larger ones before they are dragged into the central star.

This mechanism for trapping particles within the disc is an interesting alternative to streaming instabilities, the clumping of particles in a gaseous disc with pressure support \citep[e.g.,][]{gp00, yg05, jy07, yj07, johansen09}. In streaming instabilities, efficient clumping requires dust-to-gas ratios of order unity near the disc midplane. Vortices require no special dust-to-gas ratio.  \emph{Hence, vortices above the disc midplane and streaming instabilities in the midplane may provide complementary mechanisms for forming large planetesimals from small particles.}

To examine whether vortices can trap particles in protoplanetary discs, we consider several accretion disc models.  To develop an initial picture of this process, we start with a standard, static protostellar disc where the surface density and temperature decrease radially outward.  To explore a broad range of discs, we consider masses and surface density gradients consistent with observations of discs surrounding young stars.  Our results suggest that four vortex zones generically exist in discs with surface density and temperature profiles that are decreasing, power-law functions of $r$.  Two of these vortex zones form an annulus at intermediate distances from the star, within which particles are tightly coupled to the gas and are considered ``small.'' Vortex capture of particles is optimal near and at the boundaries of this annulus.  To explore the generality of these results, we then consider two models of evolving discs: a completely viscous disc (\S\ref{sect:viscous}) and an irradiated, viscous disc (\S\ref{sect:irradiated}).  Our analyses suggests that these basic conclusions are unaffected by the choice of disc model.

\section{Vortices in a Static, MMSN Disc}
\label{sect:static}

Determining the locations where vortices trap particles in gaseous discs involves specifying the surface density and temperature profiles.  In models of the minimum mass solar nebula (MMSN; \citealt{weiden77b,hayashi81,cuzzi93}),
\begin{equation}
\begin{split}
&\Sigma = \Sigma_0 \left( \frac{r}{r_0} \right)^{-p},\\
&T = T_0 \left( \frac{r}{r_0} \right)^{-q}.\\
\end{split}
\label{eq:profiles_mmsn}
\end{equation}
At $r_0 = 1$ AU, the normalization values typically chosen are $\Sigma_0 = 1700$ g cm$^{-2}$ and $T_0 = 280$ K.  In this section, we assume $\gamma=1$, i.e., the sound speed is isothermal.  Standard choices for the surface density and temperature indices are $1/2 \le q \le 3/4$ and $0 \le p \le 5/3$ \citep{cuzzi93}; \cite{chav00} chose $p = 3/2$ and $q = 1/2$.  For $p=3/2$, evaluating $2\pi \int \Sigma rdr$ from the approximate locations of Mercury to Neptune (0.22--35.5 AU) yields a total mass of about $0.01~M_\odot$ \citep{weiden77b}.  Repeating the same exercise with $p=0$ yields a mass of about $0.8~M_\odot$.  

Submillimetre studies of young stellar objects show that the surface density drops off more slowly than predicted by the $p=3/2$ MMSN model, and that the distribution of disc masses is roughly log-normal with a typical mass $\sim 0.01~M_\odot$ \citep{aw05,aw07}.  Detection of continuum emission at 1.3 mm from 11 circumstellar discs around low- and intermediate-mass pre-main-sequence stars yields the constraint $-0.8 \lesssim p \lesssim 0.8$ \citep{isella09}; the disc mass ranges from $0.07~M_\odot$ ($p=0.8$) to $9~M_\odot$ ($p=-0.8$) if we keep $\Sigma_0 = 1700$ g cm$^{-2}$.  In this ensemble, the most massive discs ($\sim 0.1 ~M_\odot$) can produce giant planets by either core accretion \citep{il04} or gravitational instability \citep{rafikov09,kratter10,mb10}.  Lower mass discs may produce Pluto-mass objects and debris discs \citep{ht10,kb10}.  Inferred disc outer radii from partially-resolved submillimetre images of these $\sim 1$ Myr-old objects are typically $\sim 200$ AU; a viscous disc model with $\alpha \sim 0.01$ matches the median spectral energy and surface brightness distributions \citep{aw07}.  Many discs with $p < 0$ are dynamically unstable and may represent transient states that are less relevant for the vortex capture of particles, so we will focus on $p \ge 0$ discs.

Specifying the surface density and temperature profiles yields the vertical scale height and mean free path,
\begin{equation}
\begin{split}
&H = H_0 \left(\frac{r}{r_0} \right)^{3/2 - q/2},\\
&\lambda = \frac{2 m_{\rm H} H}{\sigma \Sigma} = \lambda_0 \left(\frac{r}{r_0}\right)^{3/2 + p -q/2},\\
\end{split}
\end{equation}
where
\begin{equation}
\lambda_0 \equiv \frac{2 m_{\rm H}}{\sigma \Sigma_0} \left(\frac{k_{\rm B} T_0 r^3_0}{G M_\star m_{\rm H}} \right)^{1/2},
\end{equation}
$\sigma \approx 2 \times 10^{-15}$ cm$^2$ is the collision cross section of a hydrogen molecule and $H_0 \equiv (k_{\rm B} T_0 r^3_0/GM_\star m_{\rm H})^{1/2}$.  For $M_\star = M_\odot$, $H_0 \approx 0.05$ AU and $\lambda_0 \approx 0.8$ cm.  The mass density is then $\rho_0 \sim \Sigma_0/H_0 \sim 10^{-9}$ g cm$^{-3}$.  It follows that the pressure scales as $P \propto r^{-(p+q/2+3/2)}$.  Inward radial drift occurs when the pressure decreases outward, which requires (equation [\ref{eq:dv}])
\begin{equation}
p > - \frac{\left( q + 3 \right)}{2}.
\end{equation}
For $q=1/2$, we need $p > -7/4$ for outward radial drift.

\begin{figure}
\begin{center}
\includegraphics[width=0.5\columnwidth]{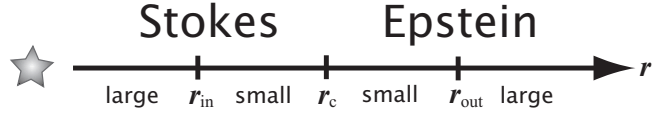}
\end{center}
\vspace{-0.2in}
\caption{Schematic of different vortex zones within a gaseous disc.  Small and large particles have $\xi/\Omega \gg 1$ and $\xi/\Omega \ll 1$, respectively (see text).}
\label{fig:schematic}
\end{figure}

\cite{chav00} realized that there are four vortex zones within a gaseous disc (Figure \ref{fig:schematic}).  Beyond a critical distance $r=r_{\rm c}$, there is a transition from the Stokes ($a > 9\lambda/4$) to the Epstein ($a < 9\lambda/4$) regimes.  The critical distance is bounded by two other transitional distances $r_{\rm in}$ and $r_{\rm out}$.  The annulus $r_{\rm in} \le r \le r_{\rm out}$ defines a region within which particles of a given size are tightly coupled to the gas.  It is conceivable that coagulation takes place in such conditions \citep[e.g.,][]{inaba06}.  At the boundaries of the annulus ($r=r_{\rm in}$ and $r=r_{\rm out}$), particles are captured by vortices within an orbital period ($t_{\rm cap} \sim \Omega^{-1}$).  Outside of the annulus ($r < r_{\rm in}$ and $r > r_{\rm out}$), capture times become long and radial drift dominates.  If particles grow rapidly within vortices, then $r=r_{\rm in}$ and $r=r_{\rm out}$ are natural locations for planet formation. Indeed, \cite{chav00} adopted a MMSN model and concluded that the Earth and Jupiter could have formed in-situ via core accretion at these locations.  In a time-dependent disc, however, the locations and sizes of the vortex zones illustrated in Figure \ref{fig:schematic} evolve with time. Coagulation and vortex capture then need to work in tandem to concentrate particles successfully.  Before we embark on this task, we briefly review and generalize the static, MMSN results of \cite{chav00}.  

Upon specifying the disc model (equation [\ref{eq:profiles_mmsn}]), we can compute the critical distance that separates the Epstein and Stokes regimes,
\begin{equation}
r_{\rm c} = \left(\frac{4a}{9\lambda_0} \right)^{2/\left( 3 + 2p - q \right)} r_0.
\label{eq:critical}
\end{equation}
For $a \sim \lambda_0 \sim 1$ cm, $r_{\rm c} \sim 1$ AU.  In each regime, $\xi/\Omega > 1$ (small particles) when $r_{\rm in} < r < r_c$ and $r_c < r < r_{\rm out}$,
\begin{equation}
\begin{split}
&r_{\rm in} = r_0 \left( \frac{8 \rho_s a^2}{9 \Sigma_0 \lambda_0} \right)^{2/\left(3-q\right)}, \\
&r_{\rm out} = r_0 \left( \frac{\Sigma_0}{2 \rho_s a} \right)^{1/p}. \\
\end{split}
\end{equation}
Effectively, the disc can be divided into 4 sub-regions or zones as illustrated in Figure \ref{fig:schematic}.  Two of the zones form an annulus within which particles are considered small, even if the entire disc is populated with monodisperse particles (i.e., particles of the same size).  Each zone has a different expression for the capture time:
\begin{equation}
\Omega t_{\rm cap} = 
\begin{cases}
\frac{64 \rho_s a^2}{27 \Sigma_0 \lambda_0} \left(\frac{r}{r_0} \right)^{-\left(3-q\right)/2}, & r < r_{\rm in}, \\
\frac{3 \Sigma_0 \lambda_0}{\rho_s a^2} \left(\frac{r}{r_0} \right)^{\left(3-q\right)/2}, & r_{\rm in} < r < r_c, \\
\frac{4 \Sigma_0}{3 \rho_s a} \left(\frac{r}{r_0} \right)^{-p}, & r_c < r < r_{\rm out}, \\
\frac{16 \rho_s a}{3 \Sigma_0} \left(\frac{r}{r_0} \right)^p, & r > r_{\rm out}. \\
\end{cases}
\end{equation}
In our idealized analysis, vortex capture of particles is optimal only at $r_{\rm in}$ and $r_{\rm out}$.  In reality, there will be a small range of distances around the transitional distances where vortex capture will still occur in about a dynamical time.  We will see in \S\ref{sect:viscous} that the boundaries of the annulus ($r_{\rm in} \le r \le r_{\rm out}$) evolve with time --- specifically, the annulus shrinks as the disc ages.  In \S\ref{sect:irradiated}, the picture is complicated by the introduction of an additional transitional distance beyond which disc heating is dominated by stellar irradiation.

\begin{figure}
\begin{center}
\includegraphics[width=0.7\columnwidth]{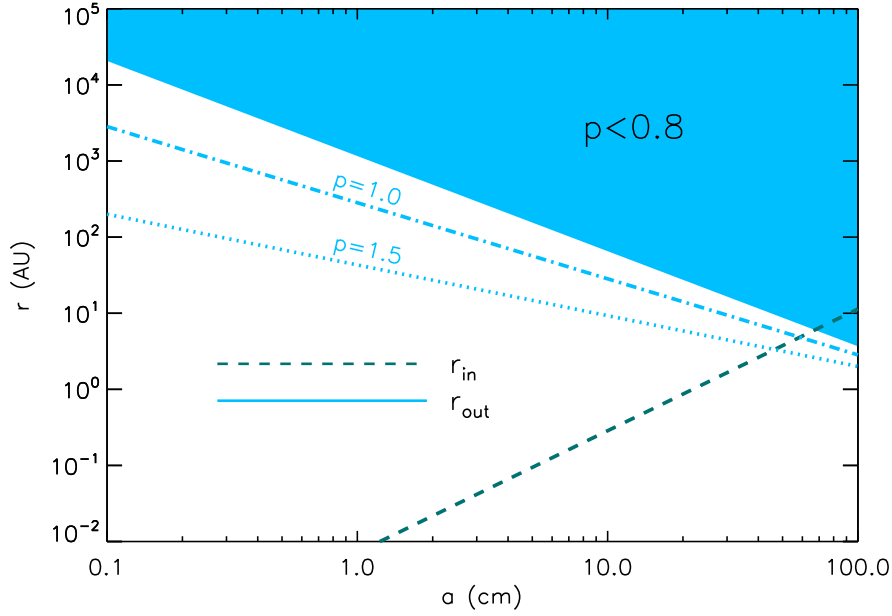}
\end{center}
\vspace{-0.2in}
\caption{Transitional distances $r_{\rm in}$ and $r_{\rm out}$ for a static, MMSN disc with $q=1/2$ as a function of particle radius.  The region with $p \le 0.8$ is shaded, while the dotted and dot-dashed lines are for $p=3/2$ and $p=1$, respectively.  The surface density and temperature normalization values are kept fixed at $\Sigma_0 = 1700$ g cm$^{-2}$ and $T_0 = 280$ K, respectively.}
\label{fig:mmsn}
\end{figure}

The allowed ranges in the indices $p$ and $q$ imply a range of values in the surface density ($\Sigma_0$) and temperature ($T_0$) normalizations, but for clarity we keep the values previously described.  We then plot $r_{\rm in}$ and $r_{\rm out}$ as functions of $a$ in Figure \ref{fig:mmsn}.   For clarity, we do not show $r_{\rm c}$.  The inner and outer transitional distances are independent of $p$ and $q$, respectively.  Since discs typically have a small range of values for $q$, $r_{\rm in}$ for a given $a$ is effectively constant for all discs; we keep $q$ fixed at 1/2.  As $p$ decreases from 3/2, the outer transitional distance moves out for a fixed $a$.  For each particle size, there is a set $\{r_{\rm in}, r_{\rm c}, r_{\rm out} \}$.  For all $a$ values, there is a locus of sets for each combination of $p$ and $q$.  For illustration, we show $r_{\rm out}$ for $p=3/2$ and $p=1$; we also shade the region for which $p \le 0.8$.  Decreasing the disc mass has no effect on $r_{\rm in}$, but causes $r_{\rm out}$ to move inward. Thus, there is some value of $a$ for which $r_{\rm in} = r_{\rm out}$.

Particle capture via vortices is only possible in discs where $r_{\rm out} > r_{\rm in}$, which allows us to derive a maximum value for the particle radius,
\begin{equation}
a < \left[ \left(\frac{9}{4 \sigma}\right)^2 \frac{k_{\rm B} T_0 r^3_0 m_{\rm H}}{GM_\star} \right]^{p/\left(3+4p-q\right)} \left( \frac{\Sigma_0}{2} \right)^{\left( 3-q\right)/\left(3+4p-q \right)} \rho_s^{-\left(3+2p-q\right)/\left(3+4p-q\right)},
\label{eq:amax2}
\end{equation}
from which specializing to $p=3/2$, $q=1/2$ yields
\begin{equation}
a < 46 \mbox{ cm} \left( \frac{T_0}{280 \mbox{ K}} \right)^{3/17} \left( \frac{M_\star}{M_\odot} \right)^{-3/17} \left( \frac{\Sigma_0}{1700 \mbox{ g cm}^{-2}} \right)^{5/17}  \left( \frac{\rho_s}{3 \mbox{ g cm}^{-3}} \right)^{-11/17}.
\end{equation}
If less massive discs are considered, the largest particle that can be captured by vortices becomes smaller.  For $p=0$, $r_{\rm out}$ is undefined.

We demand that the particles settle to the midplane of the disc by a time
\begin{equation}
\epsilon ~t_{\rm settle} < t_{\rm max},
\label{eq:effective_settle}
\end{equation}
where the dimensionless factor $\epsilon \ge 1$ accounts for an increase in the effective settling time due to vertical mixing.  This constraint yields a minimum value for the particle radius,
\begin{equation}
\begin{split}
a &> \frac{\Sigma_0 \epsilon}{\rho_s t_{\rm max}} \left(\frac{r^3_0}{GM_\star} \right)^{1/2} \left( \frac{r}{r_0} \right)^{3/2-p} \\
&> 9 \mbox{ cm } ~\epsilon \left( \frac{r}{r_0} \right)^{3/2-p} ~\left( \frac{M_\star}{M_\odot} \right)^{-1/2} \left( \frac{\Sigma_0}{1700 \mbox{ g cm}^{-2}} \right)  \left( \frac{\rho_s}{3 \mbox{ g cm}^{-3}} \frac{t_{\rm max}}{10 \mbox{ yr}} \right)^{-1}. \\
\end{split}
\label{eq:amin}
\end{equation}
Equations (\ref{eq:amax2}) and (\ref{eq:amin}) collectively yield a maximum value for $\Sigma_0$ independent of $a$.  In other words, we can estimate a maximum mass for a disc in which particles can both settle and be trapped by vortices.  Let $s_0$ and $s^\prime_0$ denote the outer and inner disc radii, respectively; we further assume $s_0 \gg s^\prime_0$.  Combining equations (\ref{eq:amax2}) and (\ref{eq:amin}) yields
\begin{equation}
M_0 < 2^{\left(4p + q-3\right)/4p} \pi \left[ \left( \frac{9 \rho_s}{4 \sigma} \right)^2 k_{\rm B} T_0 m_{\rm H} \right]^{1/4} \left( \frac{GM_\star}{r^3_0} \right)^{\left(3+2p-q\right)/8p} \left(\frac{t_{\rm max}}{\epsilon}\right)^{\left(3+4p-q\right)/4p} r_0 ~\int^{s_0}_{s^\prime_0} \left(\frac{r}{r_0} \right)^{1-p-\beta_1} ~dr,
\end{equation}
where $\beta_1 \equiv (3-2p)(3+4p-q)/8p$.  Specializing to $p=3/2$ ($\beta_1=0$) and $q=1/2$ produces a more wieldy expression,
\begin{equation}
M_0 < 0.11 ~M_\odot  ~\epsilon^{-17/12} ~\left( \frac{s_0}{20 \mbox{ AU}} \frac{\rho_s}{3 \mbox{ g cm}^{-3}} \right)^{1/2} \left( \frac{T_0}{280 \mbox{ K}} \right)^{1/4} \left( \frac{M_\star}{M_\odot} \right)^{11/24} \left( \frac{t_{\rm max}}{10 \mbox{ yr}} \right)^{17/12}.
\label{eq:min_mass}
\end{equation}
This mass is comparable to the maximum disc mass observed in protostellar discs with ages of 1 Myr \citep{aw05,aw07}.  Note that there is no constraint on the disc mass when
\begin{equation}
p = \frac{3-q}{2}.
\end{equation}
For $q=1/2$, this constraint occurs when $p=5/4$.  The mass constraint on a disc hosting particles that can both settle and be captured by vortices depends somewhat weakly on the disc properties and sensitively on the maximum time imposed for settling; $t_{\rm max}$ may be interpreted as the radial drift time or even the gas dissipation time.  Conversely, equation (\ref{eq:min_mass}) informs us that there should exist many discs in which the particles may be captured by vortices before having a chance to settle to the disc midplanes.  In both scenarios, the settling/drifting particles may be sequestered by off-midplane vortices that are effectively 2D (see equation [\ref{eq:froude}]).  As demonstrated by the scaling dependence of $\epsilon$, the condition in equation (\ref{eq:min_mass}) becomes more restrictive (i.e., lower maximum mass) when vertical mixing is present.

Our static, MMSN model allows us to draw the following conclusions: in a disc where the surface density and temperature are power-law functions that decrease with radius, four vortex zones exist for $a \lesssim 50$ cm.  Two of the vortex zones form an annulus at intermediate distances where particles of a given size are considered ``small'' even if the entire disc is populated with particles of the same size.  Vortex capture occurs on a dynamical time scale near and at the boundaries of this annulus --- capture occurs for particle sizes which coagulation likely produces in discs with masses comparable to those of observed discs. We next consider an evolving disc dominated by viscous heating, where the transitional distances discussed in this section and Figure \ref{fig:schematic} are allowed to evolve.

\section{Vortices in a Viscously-Heated Disc}
\label{sect:viscous}

\cite{cham09} considers an evolving disc with a structure determined solely by viscous heating. He solves equations for the vertical structure and energy balance to derive analytic expressions for the disc mass, radius, surface density, and temperature as functions of time. The surface density and temperature are power-laws in radial distance and time,
\begin{equation}
\begin{split}
&\Sigma = \Sigma_{\rm vis} \left(\frac{r}{s_0}\right)^{-3/5} \left( 1 + \tau \right)^{-57/80},\\
&T = T_{\rm vis} \left(\frac{8}{3 \kappa_0 \Sigma_{\rm vis}} \right)^{1/4} \left( \frac{r}{s_0} \right)^{-3/4} \left( 1 + \tau \right)^{-19/64},\\
\end{split}
\label{eq:vismodel}
\end{equation}
where $s_0$ is now the initial value of the outer edge of the disc, $\kappa_0$ is the opacity of the particles (assumed constant) and $\tau \equiv t/t_{\rm vis}$.  The coefficients $\Sigma_{\rm vis}$ and $T_{\rm vis}$ are functions of the initial disc mass and outer radius. We adopt $s_0 = 20$ AU as a reasonable starting point.  We note that the models of \cite{cham09} do not consider surface density enhancements at the snow line.

The viscous time scale in the \cite{cham09} model is
\begin{equation}
t_{\rm vis} = \frac{\mu m_{\rm H} M_0 \Omega_0}{16 \pi \alpha \gamma k_{\rm B} T_{\rm vis} \Sigma_{\rm vis}},
\label{eq:viscoustime}
\end{equation}
where $\Omega_0 \equiv (GM_\star/s^3_0)^{1/2}$, $\mu$ is the mean molecular weight of the gas and $M_0$ is now the initial disc mass. Using the parameter values listed in Table \ref{tab:parameters}, $t_{\rm vis} \approx 4.4 \times 10^4$ yr.  The normalization values\footnote{Equation (20) of \cite{cham09} has a typographical error in the $\Omega^{1/3}_0$ term.} are 
\begin{equation}
\begin{split}
&\Sigma_{\rm vis} = \frac{7M_0}{10 \pi s^2_0},\\
&T_{\rm vis} = \left( \frac{27 \kappa_0 \alpha \gamma k_{\rm B} \Omega_0 \Sigma^2_{\rm vis}}{64 \sigma_{\rm SB} \mu m_{\rm H}} \right)^{1/3},\\
\end{split}
\end{equation}
where $\sigma_{\rm SB}$ is the Stefan-Boltzmann constant.  Using Table \ref{tab:parameters}, $\Sigma_{\rm vis} \approx 50$ g cm$^{-2}$ and $T_{\rm vis} \approx 27$ K.

From equation (\ref{eq:vismodel}), the vertical disc height and gas mean free path are:
\begin{equation}
\begin{split}
&H =  H_0 \left(\frac{r}{s_0} \right)^{9/8} \left( 1 + \tau \right)^{-19/128},\\
&\lambda = \frac{2 m_{\rm H} H_0}{\sigma \Sigma_{\rm vis}} \left(\frac{r}{s_0} \right)^{69/40} \left( 1 + \tau \right)^{361/640},\\
\end{split}
\label{eq:height_v}
\end{equation}
where
\begin{equation}
H_0 \equiv \left( \frac{\gamma k_{\rm B} T_{\rm vis}}{m_{\rm H}} \right)^{1/2} \left(\frac{8}{3 \kappa_0 \Sigma_{\rm vis}} \right)^{1/8} \Omega^{-1}_0.
\end{equation}

The three transitional distances illustrated in Figure \ref{fig:schematic} are
\begin{equation}
\begin{split}
&r_{\rm in} = s_0 \left(\frac{4\rho_s \sigma a^2}{9 m_{\rm H} H_0} \right)^{8/9} \left( 1 + \tau \right)^{19/144},\\
&r_{\rm c} = s_0 \left(\frac{2a \sigma \Sigma_{\rm vis}}{9 m_{\rm H} H_0} \right)^{40/69} \left( 1 + \tau \right)^{-361/1104},\\
&r_{\rm out} = s_0 \left(\frac{\Sigma_{\rm vis}}{2 \rho_s a} \right)^{5/3} \left( 1 + \tau \right)^{-19/16}.\\
\end{split}
\end{equation}
For convenience, we note that the temporal indices are approximately 0.13, -0.33 and -1.19.  In an evolving disc, the positions of the four vortex zones change with time.  The radius $r_{\rm in}$ grows with time; the radii $r_{\rm c}$ and $r_{\rm out}$ shrink with time. Thus, the region where particles of a given size are considered small ($r_{\rm in} \le r \le r_{\rm out}$) shrinks with time.  Conversely, this region is larger for smaller particles at a given moment in time.

\begin{figure}
\begin{center}
\includegraphics[width=0.48\columnwidth]{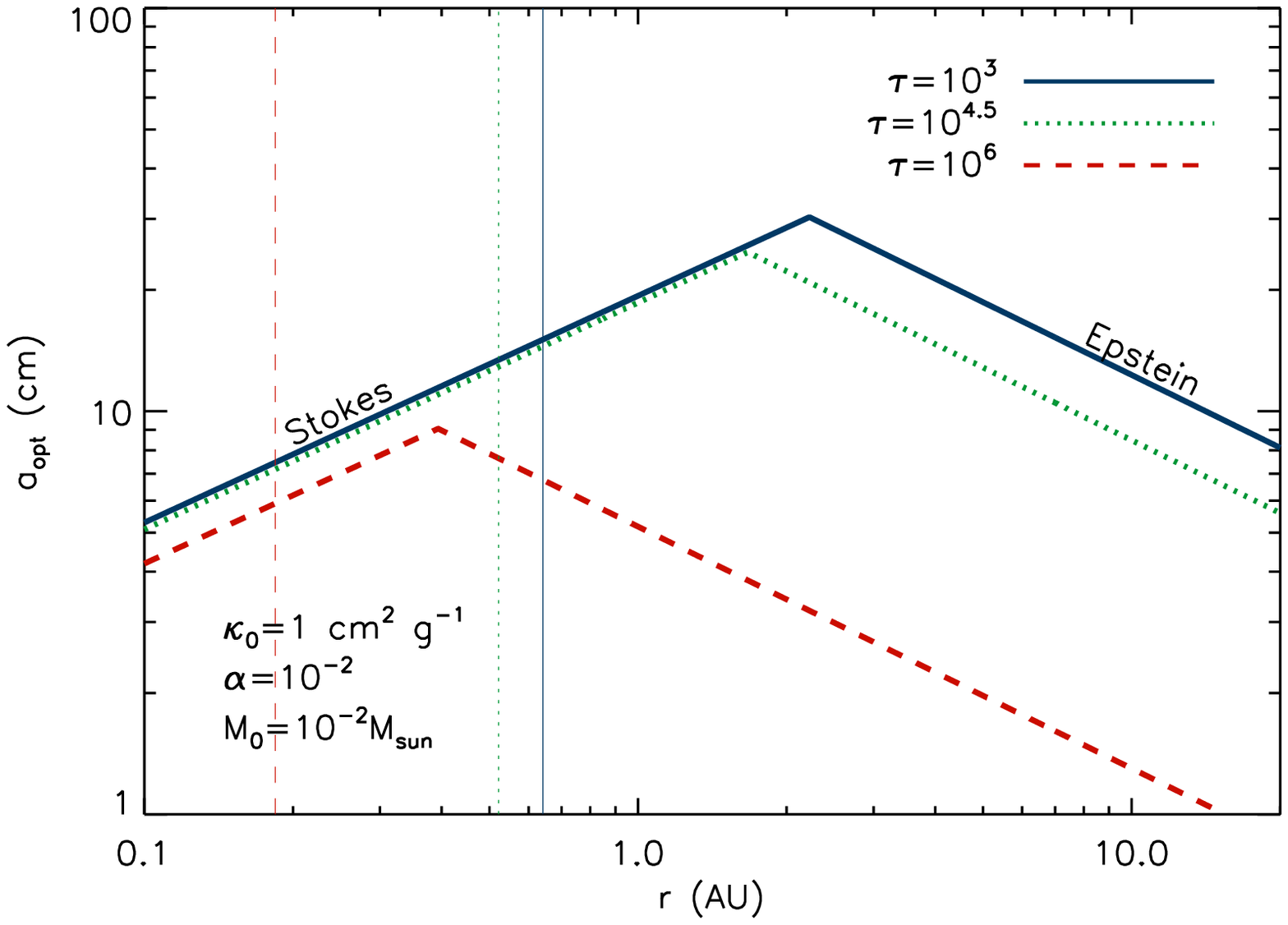}
\includegraphics[width=0.48\columnwidth]{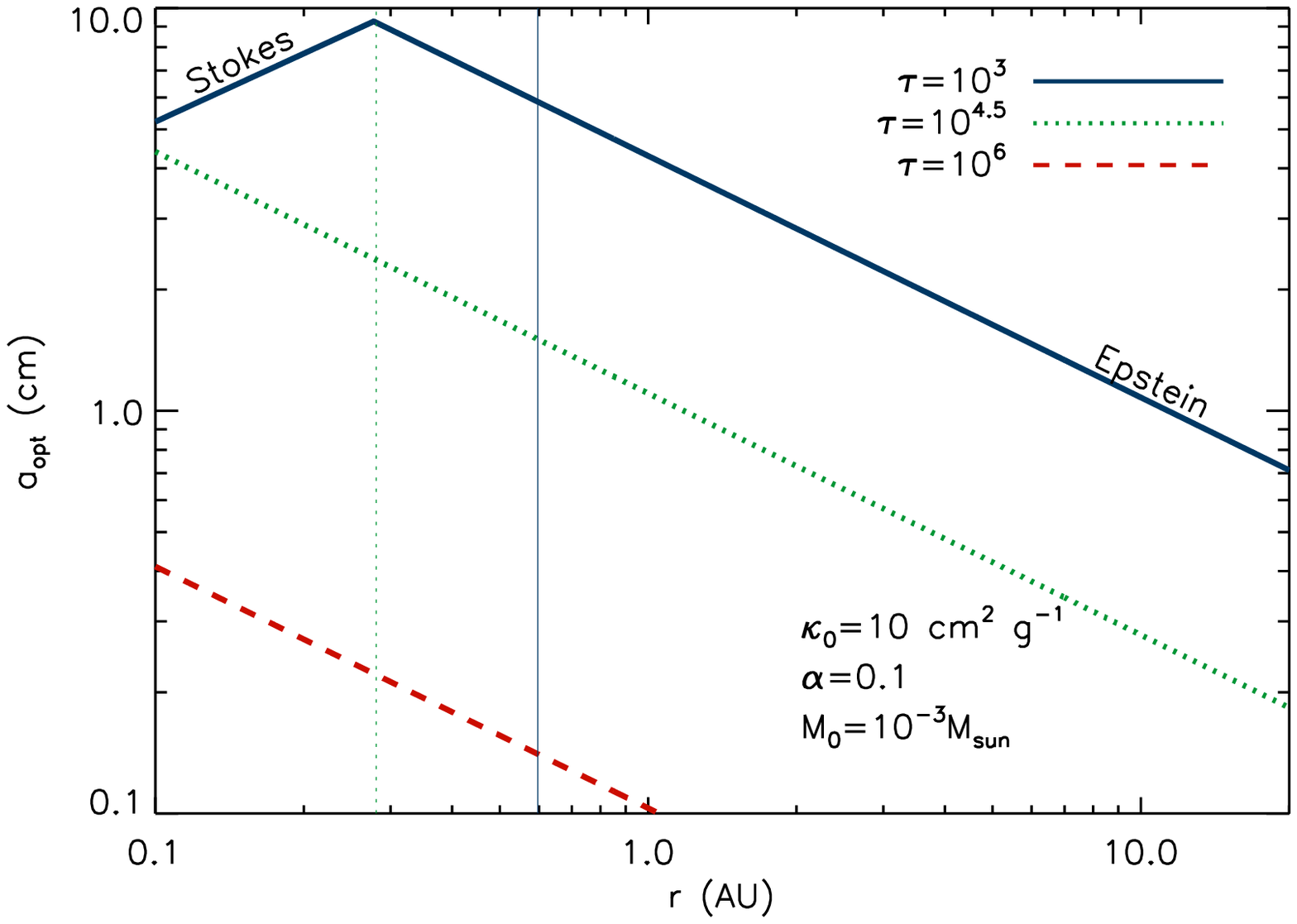}
\end{center}
\vspace{-0.2in}
\caption{Optimal radius of particles captured by vortices for evolving, viscous discs at $t=10^3, 10^{4.5}$ and $10^6$ yrs.  The vertical lines are the respective snow lines ($T=170$ K).  Left panel: using the fiducial disc parameters (Table \ref{tab:parameters}).  Right panel: a disc that is more viscous, more dust-opaque and less massive than the fiducial case.  Note that the vertical scales for the plots are different.}
\label{fig:vis}
\end{figure}

Particles with $\xi/\Omega = 1$ are preferentially captured by the vortices and by definition have $\Omega t_{\rm cap} \sim 1$ (equation \ref{eq:tcap}).  These particles have radii of
\begin{equation}
a_{\rm opt} = 
\begin{cases}
\left(\frac{9 m_{\rm H} H_0}{4 \rho_s \sigma}\right)^{1/2} \left(\frac{r}{s_0} \right)^{9/16} \left(1 + \tau \right)^{-19/256} & \mbox{(Stokes)},\\
\frac{\Sigma_{\rm vis}}{2 \rho_s} \left(\frac{r}{s_0} \right)^{-3/5} \left(1 + \tau \right)^{-57/80} & \mbox{(Epstein)}.\\
\end{cases}
\label{eq:sizes0}
\end{equation}
Again, for convenience we note that the temporal indices are approximately -0.074 and -0.71.  

\emph{As the disc ages, the vortices prefer to pick out smaller particles, a phenomenon we term ``vortex aging''.}  In Figure \ref{fig:vis}, we plot $a_{\rm opt}$ as a function of $r$ at $t=10^3, 10^{4.5}$ and $10^6$ yrs, representing the possible range of coagulation time scales inferred from both theory and observations \citep{bw08,cy10}.  If our fiducial, evolving, viscous disc is capable of generating vortices, these vortices prefer capturing particles with radii $\sim 1$--10 cm (left panel of Figure \ref{fig:vis}).  If coagulation builds particles up to maximum sizes $\sim 1$ mm, then at $t=10^6$ yrs vortex capture is only possible at $r \gtrsim 16$ AU.  

Analogous to equation (\ref{eq:amax2}) in \S\ref{sect:static}, the maximum particle radius imposed by the condition $r_{\rm out} > r_{\rm in}$ is

\begin{equation}
\begin{split}
&a < 31 \mbox{ cm} ~\left(\frac{M_\star}{M_\odot} \right)^{-10/93} \left(\frac{s_0}{20 \mbox{ AU}} \right)^{10/31} \left(\frac{\Sigma_{\rm vis}}{50 \mbox{ g cm}^{-2}} \right)^{50/93} \left(\frac{\mu}{2.4} \right)^{-4/93} \left(\frac{\gamma}{1.4} \right)^{16/93} \left(\frac{\alpha}{0.01} \right)^{4/93} \\
& ~\times \left(\frac{\rho_s}{3 \mbox{ g cm}^{-3}} \right)^{-23/31} \left(\frac{\kappa_0}{1 \mbox{ cm}^2\mbox{ g}^{-1}} \right)^{1/93} \left( 1 + \tau \right)^{-95/248}.
\end{split}
\label{eq:amax_viscous}
\end{equation}
The minimum particle radius set by equation (\ref{eq:effective_settle}) is
\begin{equation}
\begin{split}
a &> \frac{\Sigma_{\rm vis} \epsilon}{\rho_s \Omega_0 t_{\rm max}} \left(\frac{r}{s_0} \right)^{9/10} \left( 1+ \tau \right)^{-57/80} \\
&> 24 \mbox{ cm } ~\epsilon \left(\frac{r}{s_0} \right)^{9/10} ~\left(\frac{M_\star}{M_\odot} \right)^{-1/2} \left(\frac{s_0}{20 \mbox{ AU}} \right)^{3/2} \left(\frac{\Sigma_{\rm vis}}{50 \mbox{ g cm}^{-2}} \right) \left(\frac{\rho_s}{3 \mbox{ g cm}^{-3}} \frac{t_{\rm max}}{10 \mbox{ yr}} \right)^{-1} \left( 1+ \tau \right)^{-57/80}. \\
\end{split}
\label{eq:amin_viscous}
\end{equation}

By combining equations (\ref{eq:amax_viscous}) and (\ref{eq:amin_viscous}), it follows that discs which are able to accommodate both particle settling and vortex capture are subjected to a constraint on their masses,
\begin{equation}
\begin{split}
& M < 0.28 ~M_\odot ~\epsilon^{-93/43} ~\left(\frac{s_0/s^\prime_0}{100} \right)^{47/86} \left(\frac{M_\star}{M_\odot} \right)^{73/86} \left(\frac{s_0}{20 \mbox{ AU}} \right)^{-47/86} \left(\frac{\mu}{2.4} \right)^{-4/43} \left(\frac{\gamma}{1.4} \right)^{16/43} \left(\frac{\alpha}{0.01} \right)^{4/43} \\
& ~\times \left(\frac{\rho_s}{3 \mbox{ g cm}^{-3}} \right)^{24/43} \left(\frac{\kappa_0}{1 \mbox{ cm}^2\mbox{ g}^{-1}} \right)^{1/43} \left(\frac{t_{\rm max}}{10 \mbox{ yr}} \right)^{93/43} \left( 1 + \tau \right)^{1767/2480}.
\end{split}
\label{eq:maxmass_viscous}
\end{equation}
For convenience, we note that the temporal index is approximately 0.71.  As in \S3, the disc mass constraint depends weakly on the stellar and disc properties, but it is sensitive to the maximum time imposed for settling. For likely ranges of disc parameters, the maximum disc mass is consistent with the masses of observed discs.

For vortices in discs to capture particles, they need to age on a time scale comparable to that required for coagulation, and to attain capture radii comparable to the maximum particle size attainable via coagulation.  Lower values of $a_{\rm opt}$ are obtained when discs are generally more viscous, more dust-opaque or less massive.  We have examined each of these cases separately, but do not show them; instead, we plot the extreme example of a more viscous, more dust-opaque \emph{and} less massive disc in the right panel of Figure \ref{fig:vis}.  Note that the vertical scales in the left and right panels are different.  In this case, it is clear that $a_{\rm opt} \sim 1$ mm to $\sim 10$ cm for $t \sim 10^3$--$10^6$ yrs.  Almost the entire disc is in the Epstein regime, where the vortices age more rapidly.  Our results may be invalid when $r \ll 1$ AU because non-linear Stokes drag becomes important, which occurs when the gas Reynolds number,
\begin{equation}
{\cal R} \sim \frac{2a \sigma \Sigma_{\rm vis}}{m_{\rm H}} s^{3/5}_0 r^{-8/5} \left( 1 + \tau \right)^{-57/80},
\end{equation}
greatly exceeds unity.  It follows that non-linear Stokes drag sets in when
\begin{equation}
r \ll 0.15 \mbox{ AU} ~\left( \frac{a}{1 \mbox{ cm}} \frac{\Sigma_{\rm vis}}{50 \mbox{ g cm}^{-2}} \right)^{5/8} \left(\frac{s_0}{20 \mbox{ AU}} \right)^{3/8} \left( 1 + \tau \right)^{-57/128}.
\end{equation}

\section{Vortices in a Viscous, Irradiated Disc}
\label{sect:irradiated}

\cite{cham09} also considers the case of an evolving, viscous, irradiated disc.  Whether a disc has an irradiated outer region depends upon comparing two characteristic temperatures,
\begin{equation}
\begin{split}
&T_{\rm vis} = \left( \frac{27 \kappa_0 \alpha \gamma k_{\rm B} \Omega_0}{64 \sigma_{\rm SB} \mu m_{\rm H}} \right)^{1/3} \left(\frac{7M_0}{10 \pi s^2_0} \right)^{2/3},\\
&T_{\rm rad} = T_\star \left(\frac{4}{7} \right)^{1/4} \left(\frac{T_\star}{T_c} \right)^{1/7} \left(\frac{R_\star}{s_0} \right)^{3/7},\\
\end{split}
\label{eq:temperatures}
\end{equation}
where $T_c \equiv GM_\star \mu m_{\rm H}/k_{\rm B} R_\star$, $T_\star$ is the stellar effective temperature and $R_\star$ is the stellar radius.  Using Table \ref{tab:parameters}, $T_{\rm vis} \approx 27$ K and $T_{\rm rad} \approx 38$ K; since $T_{\rm rad} > T_{\rm vis}$, the fiducial outer disc is dominated by stellar irradiation rather than viscous heating.  However, if $M_0$, $\kappa_0$ or $\alpha$ are increased by an order of magnitude, we get $T_{\rm rad} < T_{\rm vis}$ --- the entire disc is initially dominated by viscous heating.  For clarity, we first consider the case where the disc has an initial irradiated region; we subsequently consider a disc where the irradiated region develops at a later time.

\subsection{With Initial Irradiated Region ($T_{\rm rad} > T_{\rm vis}$)}
\label{subsect:initial}

The characteristic surface densities are
\begin{equation}
\begin{split}
&\Sigma_{\rm vis} = \Sigma_{\rm rad} \left(\frac{T_{\rm rad}}{T_{\rm vis}} \right)^{4/5}, \\
&\Sigma_{\rm rad} = \frac{13 M_0}{28 \pi s^2_0} \left[1 - \frac{33}{98} \left(\frac{T_{\rm vis}}{T_{\rm rad}} \right)^{52/33} \right]^{-1}. \\
\end{split}
\label{eq:densities}
\end{equation}
Using Table \ref{tab:parameters}, $\Sigma_{\rm vis} \approx 54$ g cm$^{-2}$ and $\Sigma_{\rm rad} \approx 41$ g cm$^{-2}$.  When $T_{\rm rad} > T_{\rm vis}$, we replace $\Sigma_{\rm vis}$ and $T_{\rm vis}$ in equation (\ref{eq:viscoustime}) by $\Sigma_{\rm rad}$ and $T_{\rm rad}$, respectively; our fiducial disc now has $t_{\rm vis} \approx 3.8 \times 10^4$ yrs.

The surface density profile is
\begin{equation}
\Sigma = 
\begin{cases}
\Sigma_{\rm vis} \left(\frac{r}{s_0} \right)^{-3/5} \left(1 + \tau \right)^{-57/80}, & r < r_{\rm t}, \\
\Sigma_{\rm rad} \left(\frac{r}{s_0} \right)^{-15/14} \left(1 + \tau \right)^{-19/16}, & r > r_{\rm t}. \\
\end{cases}
\end{equation}
Requiring the surface density to be continuous yields the transitional radius between the viscous and irradiated regions,
\begin{equation}
r_{\rm t} = s_0 \left(\frac{\Sigma_{\rm rad}}{\Sigma_{\rm vis}} \right)^{70/33} \left(1 + \tau \right)^{-133/132}.
\label{eq:rt}
\end{equation}
For our fiducial disc, the initial value of $r_{\rm t}$ is about 11 AU.  

This formulation leads to minor discontinuities in the radial temperature.  In the viscous regime, the disc is not vertically isothermal.  Thus, the effective temperature is lower than the midplane temperature.  In the irradiated regime, the assumption of vertical isothermality requires a discontinuity in one temperature at the boundary between the two regimes.  \cite{cham09} adopts a model where the midplane temperature,
\begin{equation}
T_{\rm mid} = T_{\rm rad}
\begin{cases}
\frac{\Sigma_{\rm rad}}{\Sigma_{\rm vis}} \left(\frac{r}{s_0} \right)^{-9/10} \left(1 + \tau \right)^{-19/40}, & r < r_{\rm t}, \\
\left(\frac{r}{s_0} \right)^{-3/7}, & r > r_{\rm t}, \\
\end{cases}
\end{equation}
is continuous by construction, but the effective temperature,
\begin{equation}
T = T_{\rm rad}
\begin{cases}
\frac{\Sigma_{\rm rad}}{\Sigma_{\rm vis}} \left( \frac{8}{3 \kappa_0 \Sigma_{\rm vis}} \right)^{1/4} \left(\frac{r}{s_0} \right)^{-3/4} \left(1 + \tau \right)^{-19/64}, & r < r_{\rm t}, \\
\left(\frac{r}{s_0} \right)^{-3/7}, & r > r_{\rm t}, \\
\end{cases}
\end{equation}
is then discontinuous at $r = r_{\rm t}$.  Despite this lack of continuity, the jump in the effective temperature at the boundary is fairly small.  Thus, our eventual estimates for $a_{\rm opt}$ near $r=r_{\rm t}$ should be correct to within a factor of a few.

Knowledge of the effective temperature allows us to derive the vertical scale height,
\begin{equation}
H = H_0
\begin{cases}
\left(\frac{r}{s_0} \right)^{9/8} \left(1 + \tau \right)^{-19/128}, & r<r_{\rm t},\\
\left(\frac{r}{s_0} \right)^{9/7}, & r>r_{\rm t},\\  
\end{cases}
\label{eq:height_r}
\end{equation}
where 
\begin{equation}
H_0 \equiv
\begin{cases}
\left(\frac{\gamma k_{\rm B} T_{\rm rad} \Sigma_{\rm rad}}{m_{\rm H} \Sigma_{\rm vis}} \right)^{1/2}  \left( \frac{8}{3 \kappa_0 \Sigma_{\rm vis}} \right)^{1/8} \Omega^{-1}_0, & r<r_{\rm t},\\
\left(\frac{k_{\rm B} T_{\rm rad}}{m_{\rm H}} \right)^{1/2} \Omega^{-1}_0, & r>r_{\rm t}. \\  
\end{cases}
\end{equation}
It follows that the gas mean free path is
\begin{equation}
\lambda = \frac{2 m_{\rm H} H_0}{\sigma}
\begin{cases}
\Sigma^{-1}_{\rm vis} \left(\frac{r}{s_0} \right)^{69/40} \left( 1+\tau \right)^{361/640}, & r<r_{\rm t},\\
\Sigma^{-1}_{\rm rad} \left(\frac{r}{s_0} \right)^{33/14} \left( 1+\tau \right)^{19/16}, & r>r_{\rm t}.\\
\end{cases}
\label{eq:mfp_r}
\end{equation}

\begin{figure}
\begin{center}
\includegraphics[width=0.7\columnwidth]{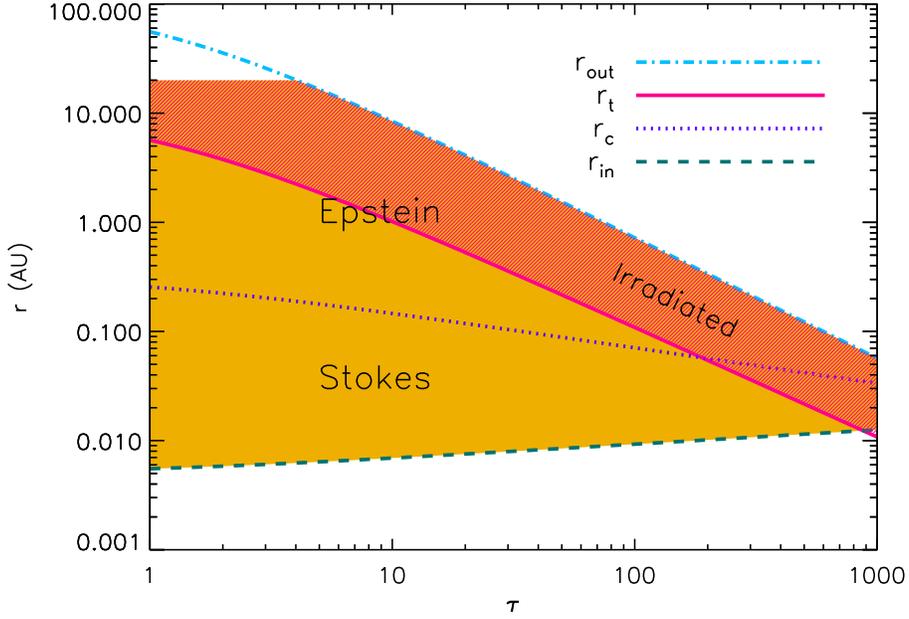}
\end{center}
\vspace{-0.2in}
\caption{Transitional distances for an evolving, viscous, irradiated disc for $a=1$ cm particles and adopting fiducial parameters (Table \ref{tab:parameters}).  The yellow shaded region is the region within which $a=1$ cm particles are considered small.  In the pink shaded region, the small particles are in the region of the disc where heating is dominated by stellar irradiation rather than viscosity.}
\label{fig:distances}
\end{figure}

The critical transitional distance between the Stokes and Epstein regions now has two expressions,
\begin{equation}
r_{\rm c} = s_0 
\begin{cases}
\frac{2 \sigma a \Sigma_{\rm vis}}{9 m_{\rm H} H_0} \left( 1 + \tau \right)^{-361/1104}, & \mbox{(viscous)},\\
\frac{2 \sigma a \Sigma_{\rm rad}}{9 m_{\rm H} H_0} \left( 1 + \tau \right)^{-133/264}, & \mbox{(irradiated)}.\\
\end{cases}
\end{equation}
For the range of parameters considered in this study, we always have $r_{\rm c} < r_{\rm t}$ at $t=0$, implying that the discs begin with $r_{\rm c}$ in the viscous regime.  Since $r_{\rm t}$ decreases more strongly with time (equation \ref{eq:rt}), it is possible for some discs with a \emph{monodisperse} population of particles to have $r_{\rm c}$ in the irradiated regime at a later stage of evolution.  For our fiducial disc, $r_{\rm c}(t=0) \approx 0.3$ AU.  The other transitional distances are
\begin{equation}
r_{\rm in} = s_0 
\begin{cases}
\left(\frac{4 \rho_s a^2 \sigma}{9 m_{\rm H} H_0}\right)^{8/9} \left( 1 + \tau \right)^{19/144}, & \mbox{(viscous)},\\
\left(\frac{4 \rho_s a^2 \sigma}{9 m_{\rm H} H_0}\right)^{7/9}, & \mbox{(irradiated)},\\
\end{cases}
\end{equation}
and
\begin{equation}
r_{\rm out} = s_0 
\begin{cases}
\left(\frac{\Sigma_{\rm vis}}{2\rho_s a} \right)^{5/3} \left( 1 + \tau \right)^{-19/16}, & \mbox{(viscous)},\\
\left(\frac{\Sigma_{\rm rad}}{2\rho_s a} \right)^{14/15} \left( 1 + \tau \right)^{-133/120}, & \mbox{(irradiated)}.\\
\end{cases}
\end{equation}
Although we have shown both cases for completeness, we typically have $r_{\rm in}(t=0)$ and $r_{\rm out}(t=0)$ in the viscous and irradiated regimes, respectively; their fiducial values are 0.005 AU and 120 AU.

Formally, there are four different scaling relations for the optimal capture radius, depending on whether $r_{\rm t} > r_{\rm c}$ or $r_{\rm t} < r_{\rm c}$:
\begin{equation}
a_{\rm opt} = 
\begin{cases}
\sqrt{\frac{9 m_{\rm H} H_0}{4 \rho_s \sigma}} \left(\frac{r}{s_0} \right)^{9/16} \left(1+\tau \right)^{-19/256}, & r < r_{\rm c} < r_{\rm t} ~\mbox{ or } ~r < r_{\rm t} < r_{\rm c}, \\
\frac{\Sigma_{\rm vis}}{2 \rho_s} \left(\frac{r}{s_0} \right)^{-3/5} \left(1+\tau \right)^{-57/80}, & r_{\rm c} < r < r_{\rm t}, \\
\sqrt{\frac{9 m_{\rm H} H_0}{4 \rho_s \sigma}} \left(\frac{r}{s_0} \right)^{9/14}, & r_{\rm t} < r < r_{\rm c}, \\
\frac{\Sigma_{\rm rad}}{2 \rho_s} \left(\frac{r}{s_0} \right)^{-15/14} \left(1+\tau \right)^{-19/16}, & r_{\rm c} < r_{\rm t} < r ~\mbox{ or } ~r_{\rm t} < r_{\rm c} < r. \\
\end{cases}
\label{eq:sizes}
\end{equation}
A generalization of the previous expressions for $a_{\rm opt}$ (equation \ref{eq:sizes0}) is that there are now three scaling relations for the optimal capture radius for any given disc.  In essence, the first and fourth equations in (\ref{eq:sizes}) are always in use; whether the second or third equation is used depends on if $r_{\rm c}$ is in the viscous or irradiated regime.

In Figure \ref{fig:distances}, we show the evolution of the four transitional distances for our fiducial disc and $a=1$ cm particles.  We have $r_{\rm out} > s_0$ initially and even after one viscous time scale, i.e., the disc has no Epstein region for large particles.  As the disc evolves, the region within which $a=1$ cm particles are considered small (i.e., they are tightly coupled to the gas) shrinks, although it does not disappear before the gas dissipates ($\tau \sim 100$--1000).  (The region bounded by $r_{\rm in} \le r \le r_{\rm out}$ is generally larger for smaller values of $a$.)  The outer disc ($r \gtrsim 11$ AU) is initially in the irradiated regime --- as the disc ages, stellar irradiation becomes increasingly important as expected.  At $\tau \approx 200$, the disc transitions from $r_{\rm c} < r_{\rm t}$ to $r_{\rm c} > r_{\rm t}$, consistent with our earlier statement about monodisperse discs evolving to having $r_{\rm c}$ in the irradiated regime.  Monodisperse discs with lower values of $M_0$ or $\alpha$ require less viscous times to evolve to the $r_{\rm c} > r_{\rm t}$ stage.  

\subsection{Irradiated Region Develops Later ($T_{\rm rad} < T_{\rm vis}$)}

Discs with $T_{\rm rad} < T_{\rm vis}$ begin their lives entirely dominated by viscous heating as described in \S\ref{sect:viscous}.  However, at a time
\begin{equation}
\tau_{\rm appear} \equiv \frac{t_{\rm appear}}{t_{\rm vis}} = \left( \frac{T_{\rm vis}}{T_{\rm rad}} \right)^{112/73} - 1,
\end{equation}
stellar irradiation in the outer disc becomes important.  Instead of $M_0$ and $s_0$, the irradiated region is initiated with
\begin{equation}
\begin{split}
&M_1 = M_0 \left(  \frac{T_{\rm rad}}{T_{\rm vis}} \right)^{21/73},\\
&s_1 = s_0 \left(  \frac{T_{\rm vis}}{T_{\rm rad}} \right)^{42/73}.\\
\end{split}
\end{equation}

Upon estimating $\tau_{\rm appear}$, $M_1$ and $s_1$, one then revises the values for the characteristic temperatures,
\begin{equation}
\begin{split}
&T_{\rm vis} = \left( \frac{27 \kappa_0 \alpha \gamma k_{\rm B} \Omega_0}{64 \sigma_{\rm SB} \mu m_{\rm H}} \right)^{1/3} \left(\frac{7M_0}{10 \pi s^2_0} \right)^{2/3} ~\longrightarrow ~\left( \frac{27 \kappa_0 \alpha \gamma k_{\rm B} \Omega_1}{64 \sigma_{\rm SB} \mu m_{\rm H}} \right)^{1/3} \left(\frac{7M_1}{10 \pi s^2_1} \right)^{2/3},\\
&T_{\rm rad} = T_\star \left(\frac{4}{7} \right)^{1/4} \left(\frac{T_\star}{T_c} \right)^{1/7} \left(\frac{R_\star}{s_0} \right)^{3/7} ~\longrightarrow ~T_\star \left(\frac{4}{7} \right)^{1/4} \left(\frac{T_\star}{T_c} \right)^{1/7} \left(\frac{R_\star}{s_1} \right)^{3/7},\\
\end{split}
\end{equation}
where $\Omega_1 \equiv (GM_\star/s^3_1)^{1/2}$.  Other quantities are similarly revised or defined:
\begin{equation}
\begin{split}
&t_{\rm vis} = \frac{\mu m_{\rm H} M_0 \Omega_0}{16 \pi \alpha \gamma k_{\rm B} T_{\rm vis} \Sigma_{\rm vis}} ~\longrightarrow ~\frac{\mu m_{\rm H} M_1 \Omega_1}{16 \pi \alpha \gamma k_{\rm B} T_{\rm rad} \Sigma_{\rm rad}} ,\\
&\Sigma_{\rm rad} = \frac{13 M_1}{28 \pi s^2_1} \left[1 - \frac{33}{98} \left(\frac{T_{\rm vis}}{T_{\rm rad}} \right)^{52/33} \right]^{-1}, \\
&\Sigma_{\rm vis} =  \frac{7M_0}{10 \pi s^2_0} ~\longrightarrow ~\Sigma_{\rm rad} \left(\frac{T_{\rm rad}}{T_{\rm vis}} \right)^{4/5}. \\
\end{split}
\end{equation}
The expressions for $H$ and $\lambda$ (equation [\ref{eq:height_v}]) are now replaced by those in equations (\ref{eq:height_r}) and (\ref{eq:mfp_r}) but with all of the revised quantities substituted.  The quantities $r_{\rm in}$, $r_{\rm c}$, $r_{\rm out}$, $r_{\rm t}$ and $a_{\rm opt}$ are also modified/introduced in a similar spirit.

\begin{figure}
\begin{center}
\includegraphics[width=0.48\columnwidth]{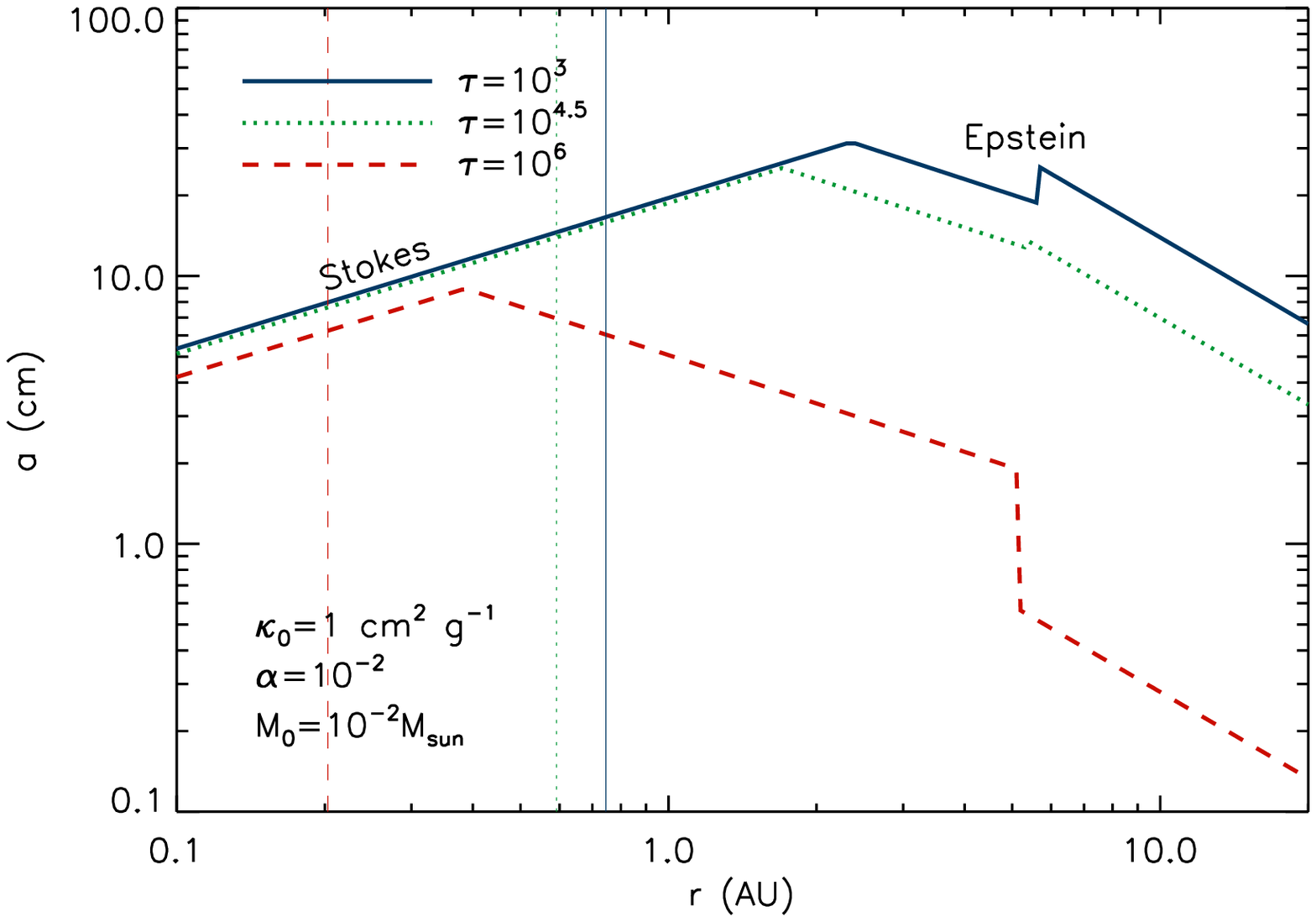}
\includegraphics[width=0.48\columnwidth]{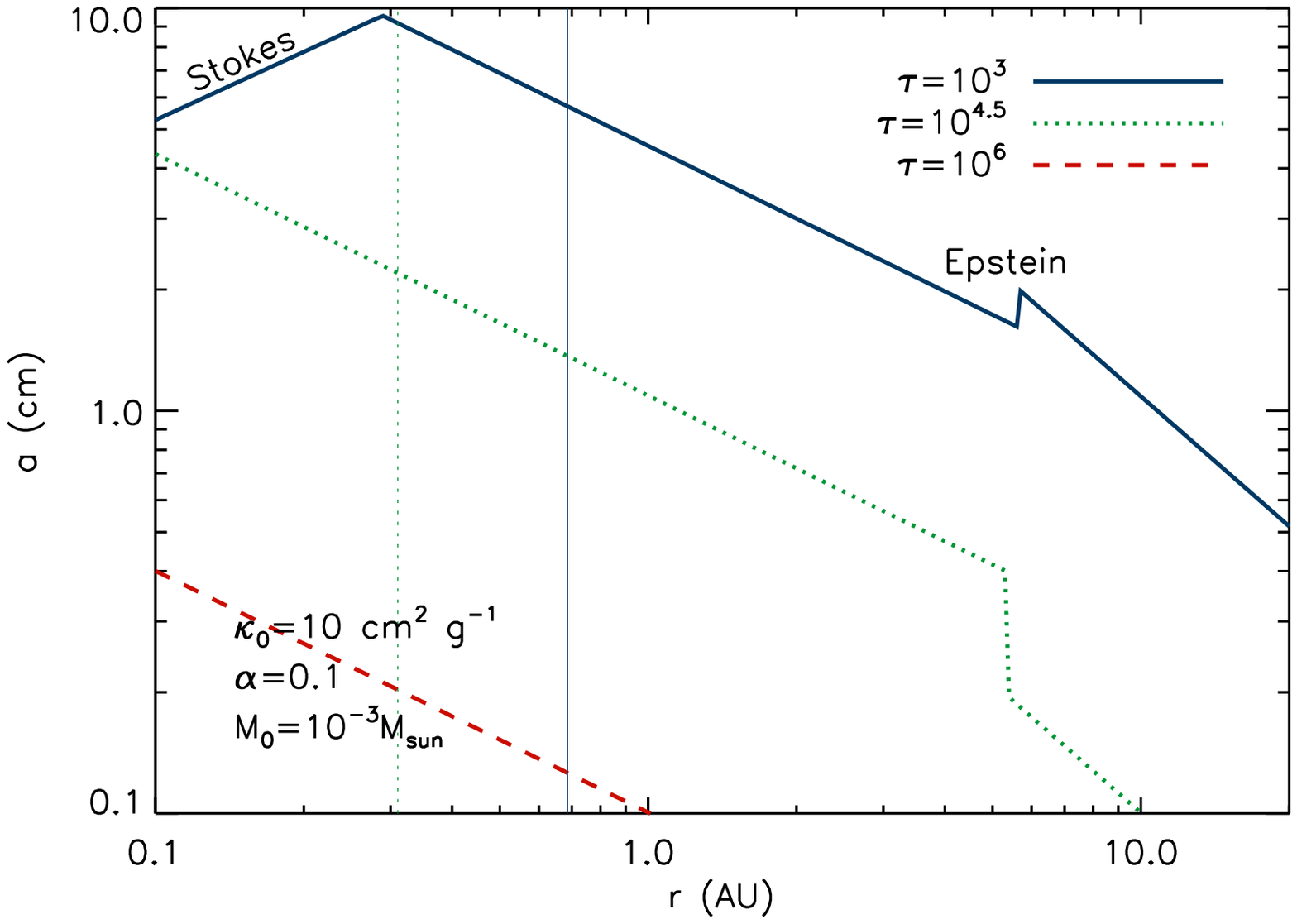}
\end{center}
\vspace{-0.2in}
\caption{Optimal radius of particles captured by vortices for evolving, viscous, irradiated discs at $t=10^3, 10^{4.5}$ and $10^6$ yrs.  The vertical lines are the respective snow lines ($T=170$ K).  Left panel: using the fiducial disc parameters (Table \ref{tab:parameters}).  Right panel: a disc that is more viscous, more dust-opaque and less massive than the fiducial case.  Note that the vertical scales for the plots are different.  The kinks in the curves are artefacts of the discontinuous nature of the effective temperature in the viscous, irradiated disc model (see text).}
\label{fig:rad}
\end{figure}

\subsection{Optimal Capture Radii}

Following \S\ref{sect:viscous} and Figure \ref{fig:vis}, we plot $a_{\rm opt}$ for evolving, viscous, irradiated discs in Figure \ref{fig:rad}.  As before, we show in the left panel the fiducial case.  The kinks in the curves are artefacts of the effective temperature profile being discontinuous at $r=r_{\rm t}$, as previously noted in \S\ref{subsect:initial}.  Nevertheless, it allows us to conclude that $1 \mbox{ mm} \lesssim a_{\rm opt} \lesssim 10$ cm, similar to our conclusions in \S\ref{sect:viscous}.  In the right panel of Figure \ref{fig:rad}, we again demonstrate that more viscous, more dust-opaque and less massive discs are more amenable to sequestering particles in vortices.  Both the discs shown in the left and right panels of Figure \ref{fig:rad} have an initial irradiated region ($T_{\rm rad} > T_{\rm vis}$).

While the exact value of $a_{\rm opt}$ at a given value of $r$ differs slightly between the viscous (\S\ref{sect:viscous}) and viscous, irradiated models, our general conclusions are the same.

\subsection{Particle Size and Disc Mass Constraints}

We generalize equations (\ref{eq:amax_viscous}), (\ref{eq:amin_viscous}) and (\ref{eq:maxmass_viscous}) to obtain constraints on viscous, irradiated discs that are capable of having particles settle to the disc midplane and also capturing them via vortices.  The minimum particle radii obtained from demanding $\epsilon ~t_{\rm settle} < t_{\rm max}$ are
\begin{equation}
a > 
\begin{cases}
\frac{\Sigma_{\rm vis} \epsilon}{\rho_s \Omega_0 t_{\rm max}} \left(\frac{r}{s_0} \right)^{9/10} \left( 1 + \tau \right)^{-57/80}, & r<r_{\rm t}, \\
\frac{\Sigma_{\rm rad} \epsilon}{\rho_s \Omega_0 t_{\rm max}} \left(\frac{r}{s_0} \right)^{3/7} \left( 1 + \tau \right)^{-19/16}, & r>r_{\rm t}, \\
\end{cases}
\label{eq:amin_rad}
\end{equation}
which can be rewritten as
\begin{equation}
a > 24 \mbox{ cm } ~\epsilon ~\left(\frac{M_\star}{M_\odot} \right)^{-1/2} \left(\frac{s_0}{20 \mbox{ AU}} \right)^{3/2} \left(\frac{\rho_s}{3 \mbox{ g cm}^{-3}} \frac{t_{\rm max}}{10 \mbox{ yr}} \right)^{-1}
\begin{cases}
\left(\frac{r}{s_0} \right)^{9/10} ~\left(\frac{\Sigma_{\rm vis}}{50 \mbox{ g cm}^{-2}} \right) \left( 1+ \tau \right)^{-57/80}, & r<r_{\rm t}, \\
\left(\frac{r}{s_0} \right)^{3/7} ~\left(\frac{\Sigma_{\rm rad}}{50 \mbox{ g cm}^{-2}} \right) \left( 1 + \tau \right)^{-19/16}, & r>r_{\rm t}. \\
\end{cases}
\label{eq:amin_rad2}
\end{equation}
For the algebra to be tractable, we make the approximation that $\Sigma_{\rm vis} \approx 7M_0/10\pi s^2_0$ only in the expression for $T_{\rm vis}$ (equation [\ref{eq:temperatures}]).  While this is formally not self-consistent (see equation [\ref{eq:densities}]), the correction made to $T_{\rm vis}$ is small.  It then follows that the maximum particle radius derived from imposing the constraint $r_{\rm out} > r_{\rm in}$ is
\begin{equation}
\begin{split}
&a < 31 \mbox{ cm} ~\left(\frac{M_\star}{M_\odot} \right)^{-376/3255} \left(\frac{s_0}{20 \mbox{ AU}} \right)^{352/1085} \left(\frac{\Sigma_{\rm vis}}{50 \mbox{ g cm}^{-2}} \right)^{242/465} \left(\frac{T_\star}{T_\odot} \right)^{32/1085} \left(\frac{R_\star}{R_\odot} \right)^{16/1085} \\
&~\times \left(\frac{\mu}{2.4} \right)^{-4/105} \left(\frac{\gamma}{1.4} \right)^{76/465} \left(\frac{\alpha}{0.01} \right)^{16/465} \left(\frac{\rho_s}{3 \mbox{ g cm}^{-3}} \right)^{-23/31} \left(\frac{\kappa_0}{1 \mbox{ cm}^2\mbox{ g}^{-1}} \right)^{1/465} \left( 1 + \tau \right)^{-95/248},\\
\end{split}
\label{eq:amax_vis2}
\end{equation}
in the viscous region of the disc ($r<r_{\rm t}$).  While the maximum radius is the same as before (equation [\ref{eq:amax_viscous}]), the scaling dependences are different because stellar irradiation modifies the temperatures and scale height.  In the irradiated region of the disc ($r>r_{\rm t}$), the maximum particle radius is
\begin{equation}
\begin{split}
&a < 41 \mbox{ cm} ~\left(\frac{M_\star}{M_\odot} \right)^{-3/35} \left(\frac{s_0}{20 \mbox{ AU}} \right)^{213/560} \left(\frac{\Sigma_{\rm vis}}{50 \mbox{ g cm}^{-2}} \right)^{23/40} \left(\frac{T_\star}{T_\odot} \right)^{-23/140} \left(\frac{R_\star}{R_\odot} \right)^{-23/280} \\
&~\times \left(\frac{\mu}{2.4} \right)^{-89/1120} \left(\frac{\gamma}{1.4} \right)^{1/10} \left(\frac{\alpha}{0.01} \right)^{1/10} \left(\frac{\rho_s}{3 \mbox{ g cm}^{-3}} \right)^{-11/16} \left(\frac{\kappa_0}{1 \mbox{ cm}^2\mbox{ g}^{-1}} \right)^{1/10} \left( 1 + \tau \right)^{-57/128}.\\
\end{split}
\label{eq:amax_rad}
\end{equation}

We next derive separate constraints on the disc mass in the viscous ($M_{\rm vis}$) and irradiated ($M_{\rm rad}$) regions of the disc.  Combining equations (\ref{eq:amin_rad}) and (\ref{eq:amax_vis2}), we get
\begin{equation}
\begin{split}
&M_{\rm vis} < 0.47~M_\odot ~\epsilon^{-465/223} ~\left(\frac{s_0/s^\prime_0}{100} \right)^{1063/2230} \left(\frac{M_\star}{M_\odot} \right)^{\beta_2} \left(\frac{s_0}{20 \mbox{ AU}} \right)^{-1409/3122} \left(\frac{T_\star}{T_\odot} \right)^{96/1561} \left(\frac{R_\star}{R_\odot} \right)^{48/1561} \\
&~\times \left(\frac{\mu}{2.4} \right)^{-124/1561} \left(\frac{\gamma}{1.4} \frac{\alpha}{0.01} \right)^{16/223} \left(\frac{\rho_s}{3 \mbox{ g cm}^{-3}} \right)^{120/223} \left(\frac{\kappa_0}{1 \mbox{ cm}^2\mbox{ g}^{-1}} \right)^{1/223} \left(\frac{t_{\rm max}}{10 \mbox{ yr}} \right)^{465/223} \left( 1 + \tau \right)^{2451/3568},\\
\end{split}
\label{eq:maxmass_vis2}
\end{equation}
where $\beta_2 = 186056/232069 \approx 0.80$ and $s^\prime_0 \ll s_0$ again denotes the inner disc radius.  For the irradiated region, combining equations (\ref{eq:amin_rad}) and (\ref{eq:amax_rad}) yields
\begin{equation}
\begin{split}
&M_{\rm rad} < 0.025~M_\odot ~\epsilon^{-8/5} ~\frac{\phi}{0.15} ~\left(\frac{M_\star}{M_\odot} \right)^{-2/7} \left(\frac{s_0}{20 \mbox{ AU}} \right)^{37/14} \left(\frac{T_\star}{T_\odot} \right)^{2/7} \left(\frac{R_\star}{R_\odot} \right)^{1/7} \\
&~\times \left(\frac{\mu}{2.4} \right)^{-1/28} \left(\frac{\rho_s}{3 \mbox{ g cm}^{-3}} \right)^{1/2} \left(\frac{t_{\rm max}}{10 \mbox{ yr}} \right)^{8/5} \left( 1 + \tau \right)^{19/16},\\
\end{split}
\label{eq:maxmass_rad}
\end{equation}
where the function $\phi$ is defined as
\begin{equation}
\phi\left(r_{\rm t}/s_0 \right) \equiv 1 - \left(r_{\rm t}/s_0 \right)^{17/70}.
\end{equation}
For $r_{\rm t}/s_0 = 0.25$--0.75, $\phi \approx 0.29$--0.07; $\phi(0.5) \approx 0.15$.  We note that $0.025~M_\odot \approx 26 M_{\rm J}$, where $M_{\rm J}$ is the mass of Jupiter.  The main differences from the viscous region are that the dependence on $s_0$ is somewhat stronger and there are no dependences on $\alpha$ or $\kappa_0$.  As expected, the disc mass in the irradiated region makes only a modest ($\sim 5\%$) contribution to the overall disc mass $M=M_{\rm vis}+M_{\rm rad}$.  In all of the models (static MMSN, viscous and viscous, irradiated discs), the maximum disc mass has strong dependences on $t_{\rm max}$ and $\epsilon$ with the power-law indices spanning about $\pm1.4$--2.2; the dependences on stellar, disc and dust properties are somewhat weak.  

We conclude that the maximum radius for particle trapping via vortices is $\sim 10$ cm, independent of disc model and with somewhat weak dependences on stellar, disc and dust properties.  We also conclude that a disc that is able to both settle particles to its midplane (in 10 yr) and capture them via vortices has a mass that is at most $\sim 0.1~M_\odot$, consistent with the masses of most observed discs.  This mass threshold decreases if vertical mixing is present in the discs to hold the particles aloft for longer than $t_{\rm settle}$.  Nevertheless, particles may drift and/or settle from off-midplane locations and be captured by vortices in about a dynamical time.

\section{Discussion}
\label{sect:discussion}

\subsection{Summary}

We have examined particle trapping by vortices in evolving, viscous and/or irradiated discs as a function of radial distance, initial disc mass, dust opacity and viscosity.  The salient points of our study are:
\begin{enumerate}

\item If the surface density and effective temperature of protoplanetary discs are decreasing, power-law functions of the distance from the star, then all discs contain four vortex zones for a fixed particle size (Figures \ref{fig:schematic} and \ref{fig:mmsn}).  

\item There is an annulus at intermediate distances from the star where small particles reside and may grow via coagulation.  The size of this annulus decreases with time and also shrinks as the particles grow (Figure \ref{fig:distances}).  Vortex capture is optimal near and at the boundaries of this annulus, and occurs within an orbital period.

\item The optimal capture radii ($\sim 1$ mm to $\sim 10$ cm; Figures \ref{fig:vis} and \ref{fig:rad}) of particles are comparable to or smaller than the sizes of particles ($\sim 1$ m) that drift fastest through the protoplanetary disc.  The enhanced particle concentration within the vortices may make the characteristic time for particle growth shorter than that for radial drift.

\item Vortices in older discs prefer to capture smaller particles, a phenomenon we term ``vortex aging'' (Figures \ref{fig:vis} and \ref{fig:rad}).  If coagulation between sub-micron particles mixed with the gas can only produce small ($\lesssim 1$ mm) particles, then vortices can capture them throughout the lifetime of the disc.  However, if coagulation manufactures larger ($\sim 10$ cm) particles when the disc is young, vortices must form early to capture them.  This metaphorical dance between coagulation and vortex aging determines how efficiently vortices help the disc to retain its mass in solids.

\item More viscous, more dust-opaque and/or less massive discs can have vortices that trap a broader range of particle sizes throughout the lifetime of the disc.  While the coagulation of grains with sizes $\lesssim 1$ mm needs to be more in synch with the evolution of more massive discs, such discs are also expected to grow grains to larger sizes more rapidly.

\item The maximum size of particle that can be trapped by vortices is $\sim 10$ cm, independent of disc model and weakly dependent on stellar, disc and dust properties.  Discs where particles settle to the midplane (within $\gtrsim 10$ yr) and are sequestered in vortices have upper limits to their masses ($\gtrsim 0.1~M_\odot$) that are consistent with those of most observed discs.  If vertical mixing is present (e.g., via turbulence), the maximum masses can be much smaller.

\end{enumerate}

\subsection{The Physics of Vortices: Open Questions}

Many open questions remain concerning the microphysics of vortices.  While we have shown that particles can be gathered by vortices, the outcome of these captures is uncertain.  Vortices may concentrate enough mass to enhance collisions rates by an order of magnitude (or more) or to trigger local gravitational instabilities \citep{aw95,gl99,gl00,kb06}. Both paths leads to the formation of self-gravitating objects which will not drift through the disc.  However, at least some of the published simulations are run in 2D (e.g., \citealt{davis00,inaba06,lyra09}) --- in the absence of viscosity or particles, such vortices live forever, implying that the mass of the planetesimal, embyro or planet formed depends either on the time the simulation is executed or the breakup of the vortex by the non-linear feedback of the concentrated particles.  It is more likely that if the centres of vortices are relatively quiescent, they then serve as nurseries for coagulation to occur between somewhat larger particles.

Vortex formation is also uncertain.  \cite{lp10} show that the subcritical baroclinic instability (SBI) is a plausible way of seeding vortices.  Discs develop this \emph{non-linear} instability when they are (radially) convectively unstable, have non-negligible thermal diffusion, and are subjected to finite vorticity perturbations ($\sim 0.1$).  The associated Reynolds number for the shearing box simulations in this study is ${\cal R} \sim 10^5$; because the threshold vorticity amplitude for invoking the SBI decreases with increasing ${\cal R}$, they speculate that the threshold amplitude could be very small (and possibly sub-sonic) in realistic discs.  

Another possibility for generating vortices is via the (linear) Rossby wave instability \citep{lovelace99,vt06}, which was invoked by \cite{inaba06} to consider 2D vortices in protoplanetary discs.  \cite{inaba06} found that the formation of vortices via the Rossby wave instability critically depends on the amplitude and width of an initial density bump placed within the disc.  This bump appears to be most unstable to perturbations with an azimuthal mode number of 5 (i.e., ``$m=5$'' perturbations).  \cite{meheut10} performed 3D simulations of stratified discs and concluded that strong and persistent vortices emerge out of the flow via the Rossby wave instability.  \cite{davis00} have noted that Rossby waves are only supported by flows with non-vanishing vorticity gradients, implying that they are relevant mostly in incompressible, inviscid flows.

The subject of vortex survivability is mired in deeper controversy.  \cite{lp09} subject vortices embedded in a shearing sheet to 3D perturbations; only vortices with $4 \lesssim q \lesssim 6$ survive the elliptical instability in unstratified discs. This stability region vanishes when stratified discs are considered.  By contrast, \cite{lithwick09} asserts that weak vortices ($q \gg 1$) can survive in quasi-2D flows.  \cite{lp10} find that vortices develop bursts of turbulence in their cores before surviving as weaker vortices; the SBI amplifies the vortices and the cycle restarts.  If turbulent vortex cores are generic and ubiquitous phenomena in realistic discs, then they may pose a setback to using vortices as mechanisms for concentrating particles.

Many collective properties of vortices in turbulent, 2D, hydrodynamic flows remain poorly understood.  Among these is the ``universal decay theory'', which is the empirical observation that the vortex density, radius, velocity, mean separation between vortices, enstrophy and kurtosis can be approximated by power laws of time parametrized by a single parameter --- a first-principles explanation is still being sought \citep{tabeling02}.  Statistical theories predict vortices to always ultimately merge, in contrast to experiments that show that above a critical separation, a pair of vortices may remain separated for many dynamical times \citep{tabeling02}.

Despite these uncertainties, vortices provide an interesting alternative to streaming instabilities \citep[e.g.,][]{gp00, yg05, jy07, yj07, johansen09}, which require dust-to-gas ratios to approach of order unity (presumably near the disc midplane).  Vortices require no special dust-to-gas ratio to trap particles and may exist off the disc midplane.  In both cases, particles with $\xi/\Omega \sim 1$ are captured and the capture size decreases as the gas in the disc dissipates.  Therefore, vortex trapping and streaming instabilities may provide complementary mechanisms for particle concentration and/or growth.  A key question to study and explore is the size distribution of planetesimals produced by each mechanism, since this might have an impact on the types of (exo)planets produced in a system.  Preliminary analyses of the initial size distribution of Solar System planetesimals suggest a broad range of values, ranging from $\sim 1$--10 km \citep{kb10} to $\sim 100$--1000 km \citep{morb09}.

\subsection{Observational Relevance}

Our study makes falsifiable predictions about the size of particles concentrated as a function of distance from the star.  Particles smaller or larger than the optimal radii for vortex capture will be uniformly distributed throughout the disc, while those with sizes $\sim 1$ mm to $\sim 10$ cm will be concentrated near their respective transitional distances $r_{\rm in}$ and $r_{\rm out}$ (Figure \ref{fig:schematic}).

If particle concentration leads to the growth of larger particles, these particles will then  decouple from the gas and migrate out of the vortices.  The decoupling will be stronger in the inner regions of the disc as $\xi/\Omega \propto a^{-2}$ (instead of $\propto a^{-1}$ in the outer regions).  In our Solar System, the smallest constituents (chondrules) of $\sim 100$ km-sized asteroids have sizes $\sim 1$ mm \citep{hewins97}, comparable to the sizes of particles trapped by vortices.  \cite{cuzzi08} (and references therein) have pointed out that there is a spread of about 1 Myr between the formation times of the oldest and youngest objects in the same meteorite, implying that particle growth was fairly inefficient.  This inference is in turn consistent with the limited temporal windows for particle trapping implied by vortex aging.  Another interesting property of $\sim 10$--100 km-sized asteroids is that many of them are formed from a physically and chemically homogeneous mix of particles of a similar size, consistent with the aerodynamic sorting property of vortices.

In extrasolar settings, observations characterizing grain opacity (and hence grain growth) in protoplanetary discs can quantify particle populations as functions of radial distance, but these are still nascent and have only been accomplished for a small number of objects, e.g., HD 163296 \citep{natta07}.  Therefore, the hypothesis that vortices serve as nurseries for particle growth will need to be tested by future infrared, submillimetre and centimetre observations of protoplanetary discs, and may lead to constraints on the time scales for grain coagulation (Figures \ref{fig:vis} and \ref{fig:rad}).  For example, the {\it Atacama Large Millimeter Array (ALMA),} which operates at wavelengths between 0.3 and 3.6 mm, might be able to detect disc features associated with vortex capture.  Coagulation-fragmentation simulations can subsequently be performed to convert the observed fluxes and $\alpha$ values into constraints on the grain properties (e.g., mass/size distribution, porosity; \citealt{birn10}).

Even if vortices do not survive long enough to create self-gravitating structures, they will certainly play a strong role in redistributing matter throughout the disc.  Whether the redistribution of matter by the vortices plays any significant role in the eventual formation of planetesimals is unknown.  An implication of a size-dependent redistribution of matter is that if the amount of electric charge carried by a particle is proportional to its size, then charge separation in protoplanetary discs may be a fairly common phenomenon.

Our study has shown that if vortices form in protoplanetary discs, they are important in discs with typical masses and for particles that are likely to condense out of the protostellar nebula.  The capture of particles also occurs at distances relevant to planet formation.  With this study, we hope to (re)ignite the debate in connecting the microphysics of vortices with the global properties of protoplanetary discs, as the first step towards understanding the \emph{efficiency} of planetesimal --- and eventually planet --- formation.

\section*{Acknowledgments}

K.H. acknowledges support from the Frank \& Peggy Taplin Membership of the Institute for Advanced Study, NASA grant NNX08AH83G and NSF grant AST-0807444.  S.K. acknowledges support from the NASA Astrophysics Theory Program, grant NNX10AF35G. We acknowledge useful discussions with Phil Armitage, Francesco Miniati, Mikhail Medvedev, Jim Stone, Jeffrey Weiss, Michael Meyer and Omar Blaes.  Improvements to the manuscript were made following reads by Anders Johansen, Andrew Youdin, John Chambers and Kevin France.  K.H. is especially grateful to Anders Johansen for several delightful conversations.  A useful report by an anonymous referee further improved the clarity of the manuscript.


\appendix
\section{Coagulation of Monomers in a Viscous, Irradiated Disc}
\label{append:coagulate}

One can estimate if binary collisions between two spherical dust grains will result in coagulation.  When two elastic spheres collide with relative velocity $v_{\rm col}$, the characteristic binding energy associated with the compressed surfaces of contact is \citep{cy10}
\begin{equation}
E_{\rm bind} \sim \delta a^2 \left(\frac{\rho_s v^2_{\rm col}}{\psi} \right)^{2/5},
\end{equation}
where $\delta$ is the surface tension from unsaturated bonds and $\psi$ is Young's modulus, a measure of the ``stiffness'' of elastic material.  Requiring the collisional energy to be less than the binding energy yields an upper limit for the relative velocity,
\begin{equation}
v_{\rm stick} \sim \left(\frac{3 \delta}{\pi a} \right)^{5/6} \rho^{-1/2}_s \psi^{-1/3}.
\end{equation}
Hence, two colliding particles will stick if $\Delta v < v_{\rm stick}$, where the velocity difference between a particle and its surrounding gas is given by equation (\ref{eq:dv}).  The maximum radii of these particles are
\begin{equation}
a_{\rm max} \sim \left( \frac{3}{\pi} \right) \left(\frac{GM_\star}{\rho_s s_0} \right)^{3/5} \delta \psi^{-2/5} T^{-6/5}_{\rm rad}
\begin{cases}
\left(\frac{80 m_{\rm H}}{99 \gamma k_{\rm B}} \right)^{6/5} \left(\frac{\Sigma_{\rm rad}}{\Sigma_{\rm vis}} \right)^{-6/5} \left(\frac{8}{3 \kappa_0 \Sigma_{\rm vis}} \right)^{-3/10} \left(\frac{r}{s_0} \right)^{3/10} \left( 1+\tau \right)^{57/160} & r<r_{\rm t}, \\
\left(\frac{28 m_{\rm H}}{39 \gamma k_{\rm B}} \right)^{6/5} \left(\frac{r}{s_0} \right)^{-3/35} & r>r_{\rm t}. \\
\end{cases}
\label{eq:amax}
\end{equation}
It is important to note that equation (\ref{eq:amax}) makes no statement about \emph{how long} it takes for coagulation to occur.  

The values to adopt for Young's modulus and the surface tension warrant some discussion.  Table 3 of \cite{chokshi93} states that $\psi = 7 \times 10^{10}$ erg cm$^{-3}$ for ice; they also list Young's modulus as $10^{11}$ erg cm$^{-3}$ for graphite and $2 \times 10^{12}$ erg cm$^{-3}$ for iron.  Terrestrial rock typically has $\psi \sim 10^{11}$--$10^{12}$ erg cm$^{-3}$.  Given the uncertainties associated with dust composition and chemistry, we consider a generous range of $\psi \sim 10^{10}$--$10^{12}$ erg cm$^{-3}$.  However, since $a_{\rm max} \propto \psi^{-2/5}$, a variation of two orders of magnitude in $\psi$ corresponds to a change of only a factor of about 6 in $a_{\rm max}$.  A somewhat larger range of uncertainty exists for the surface tension.  \cite{chokshi93} estimate $\delta = 75, 370$ and 3000 erg cm$^{-2}$ for graphite, ice and iron, respectively --- we then consider $\delta \sim 10$--$10^3$ erg cm$^{-2}$, which corresponds to two orders of magnitude of uncertainty in $a_{\rm max}$.  Collectively, we get $1.6 \times 10^{-4} \lesssim \delta \psi^{-2/5} \lesssim 0.10$.  With our fiducial disc, we find that max$\{a_{\rm max}\} \sim 0.01$--0.1 $\mu$m at $r\approx 5$--6 AU for $t\lesssim 1$ Myr.  Monomers of such sizes aggregate to form fractal dust grains, which are able grow to sizes $\sim 1$ mm because they possess internal modes of kinetic energy dissipation \citep{bw08,zsom10}.

\label{lastpage}

\end{document}